\newcommand{\non}{\nonumber}
\newcommand{\R}{{\mathbb R}}   

\def\T{\tiny\mbox{\rm T}}

\newcommand{\vectf}[4]{\left[\begin{array}{c} #1 \\  #2 \\  #3 \\ #4 \end{array}\right]}

\def\bfa{\bm a}

\def\bfe{\bm e}
\def\bff{\bm f}
\def\bfg{\bm g}

\def\bfu{\bm u}
\def\bfv{\bm v}

\def\bfx{\bm x}
\def\bfy{\bm y}
\def\bfz{\bm z}
\def\bfA{\bm A}

\def\bfC{\bm C}

\def\bfF{\bm F}

\def\bfH{\bm H}
\def\bfI{\bm I}

\def\bfP{\bm P}
\def\bfQ{\bm Q}
\def\bfR{\bm R}
\def\bfS{\bm S}

\def\bfmu{\bm\mu}

\def\bfphi{\bm\phi}

\documentclass[11pt, letterpaper]{article}

\usepackage{graphicx}
\usepackage{amsmath}
\usepackage{amssymb}
\usepackage{setspace}
\usepackage{epstopdf}
\usepackage{color}
\usepackage[letterpaper,left=1in,right=1in,top=1in,bottom=1in]{geometry}
\usepackage[linesnumbered,ruled,vlined,longend]{algorithm2e}
\usepackage{multirow}
\usepackage{longtable}
\usepackage{rotating}
\usepackage{threeparttable}
\usepackage{colortbl}
\usepackage{amsthm}
\usepackage[modulo]{lineno}
\usepackage{enumerate}
\usepackage{bm}

\newtheorem{remark}{Remark}

\title{\Large \bf Sensor network design for post-combustion CO$_2$ capture plants: Economy, complexity and robustness }

\author{\centerline{\normalsize Siyu Liu$^{a,b}$, Xunyuan Yin$^c$, Jinfeng Liu$^{b,}$\thanks{Corresponding author: J. Liu. Tel: +1-780-492-1317. Fax: +1-780-492-2881. Email: jinfeng@ualberta.ca}}\vspace{5mm}\\ 
    \centerline{\small $^{a}$ School of Internet of Things Engineering, Jiangnan University, Wuxi\ 214122, China}\\
    \centerline{\small $^{b}$ Department of Chemical \& Materials Engineering, University of Alberta,}\\
    \centerline{\small Edmonton, AB, Canada, T6G 1H9}\\
    \centerline{\small $^{c}$ School of Chemistry, Chemical Engineering and Biotechnology, Nanyang Technological University,}\\
    \centerline{\small 62 Nanyang Drive, Singapore, 637459}}
\allowdisplaybreaks

\begin{document}

\date{}

\maketitle
\setstretch{1.39}

\begin{abstract}
	State estimation is crucial for the monitoring and control of post-combustion CO$_2$ capture plants (PCCPs). The performance of state estimation is highly reliant on the configuration of sensors. In this work, we consider the problem of sensor selection for PCCPs and propose a computationally efficient method to determine an appropriate number of sensors and the corresponding placement of the sensors. The objective is to find the (near-)optimal set of sensors that provides the maximum degree of observability for state estimation while satisfying the budget constraint. Specifically, we resort to the information contained in the sensitivity matrix calculated around the operating region of a PCCP to quantify the degree of observability of the entire system corresponding to the placed sensors. The sensor selection problem is converted to an optimization problem, and is efficiently solved by a one-by-one removal approach through sensitivity analysis. Next, we extend our approach to study fault tolerance (resilience) of the selected sensors to sensor malfunction. The resilient sensor selection problem is to find a sensor network that gives good estimation performance even when some of the sensors fail, thereby improving the overall system robustness. The resilient sensor selection problem is formulated as a max-min optimization problem. We show how the proposed approach can be adapted to solve the sensor selection max-min optimization problem. By implementing the proposed approaches, the sensor network is configured for the PCCP efficiently.
\end{abstract}


\section{Introduction}

Post-combustion carbon capture using monoethanolamine (MEA) has been proven technologically viable and promising for removing carbon emissions from power plants \cite{MacDowell2010_RSC}. Effective monitoring and control of post-combustion CO$_2$ capture plants (PCCPs) is critical for safe and consistent process operation and efficient carbon capture \cite{Manaf2019_JPC}. State estimation can be exploited as the basis of both monitoring and control of PCCPs. Specifically, the estimate of the full-state is important for understanding and overseeing the status of operations, and is also essential for the controller(s) to make informed decisions on next-step process operation. The development of a new state estimation scheme for PCCP will involve two key steps: 1) to place sensors appropriately to collect real-time measurements; 2) to develop a state estimation algorithm suitable for the PCCP. Different sensor selection will lead to different sensor measurements available for state estimation, and will have significant influence on the estimation performance. In addition, the costs for sensor deployment and maintenance will also vary significantly according to sensor selection.

Sensor selection involves many factors, including location, cost, and accuracy. In general, using more sensors results in more accessible information, which helps increase the estimation accuracy and robustness. However, the number of installed sensors is typically constrained due to the limited budget. Additionally, if any sensor is installed in a way that its failure can significantly degrade measurement accuracy, it can cause issues with plant operation or even malfunction of process equipment. Therefore, it is important to develop an optimal sensor selection method that can specify the locations, number, and types of sensors while considering estimation performance and sensor cost. During PCCP operation, the system may face attacks, and some deployed sensors may malfunction. Different types of attacks have been studied previously, including denial of service (DoS) attacks \cite{Zhang2015_TAC} and false data injection attacks \cite{Mo2010_IEEECDC}. In this work, we consider scenarios where adversaries may perform DoS attacks on sensors by removing them, that is, dropping all measurement data. For this type of problem, Ye et al., showed that greedy algorithms can perform arbitrarily poorly for this problem by using a specific example \cite{Ye2020_TAC}. The graph structure of the systems was used in \cite{Ye2020_IEEETCNS} to analyze this problem.

The optimal sensor selection problem has been investigated for chemical processes \cite{LiuSY2022_ADCONIP}, water security networks \cite{Shastri2006_WRPM,Rrico2007_CCE} and other industrial applications \cite{Mkwananzi2022_JPC}. In \cite{Zhang2017_Auto}, sensor selection was addressed by minimizing the trace of the error covariance matrix produced by a Kalman filter. Another approach was employed in \cite{Alonso2004_CCE}, which used a reduced order representation of distributed process systems and determined the most appropriate sensor type and the optimal number of sensors by solving a max-min optimization problem. Greedy algorithms, which approximate the globally optimal solution to provide suboptimal results based on greedy heuristics, have also been widely used to handle sensor selection problems \cite{Clark2021_IEEESJ,Yamada2021_MSSP}. Such methods are guaranteed to perform well if the cost function is submodular \cite{Jawaid2015_Auto}. The above methods, nevertheless, are only applicable to linear systems. Recently, the sensor selection for nonlinear systems has also been considered. For example, in \cite{Yin2018_CCE}, sensor selection was conducted for wastewater treatment plants based on a reduced-order process model, yet the minimum set of sensors was established based on exhaustive search. The work in \cite{Sahoo2019_AIChE} determined the minimum number of soil moisture sensors and their optimal locations for agro-hydrological systems by using the graphical approach and the maximum multiplicity theory. 

Metrics based on the empirical observability Gramian are also commonly used for addressing sensor selection problems for nonlinear systems \cite{Singh2006_IECR,Qi2015_IEEETPS,Sumana2009_JPC}. These methods typically formulate the sensor selection problem as a mixed-integer optimization problem and resort to various existing solvers for solving the mixed-integer nonlinear programming problems, which are computationally challenging when the number of sensors is not small. Similarly, maximizing a measure of Fisher information was used to optimize discrete sensors and input designs \cite{Awasthi2020_JPC}.

Sensitivity analysis quantifies the dependence of the model outputs or states on the model parameters, which can be used for selecting the sensitive parameters for model identification \cite{Guo2021_JPC}. In our recent work \cite{LiuSY2022_DYCOPS}, we applied sensitivity analysis to simultaneous state and parameter estimation for not fully observable systems. By using the sensitivity matrix of outputs with respect to the initial state, the local observability of the state can be measured for nonlinear systems. Therefore, a computationally efficient approach to sensor selection was proposed for a wastewater treatment process exploiting the information in the sensitivity matrix of the selected sensors to the states \cite{LiuSY2022_ADCONIP}. While maximizing the degree of observability in sensor selection is crucial for estimation and control, this approach did not take into account the budget constraints for sensor deployment. Consequently, it is possible that the recommended sensors may be expensive or unaffordable to install or maintain. 

In this work, we address the problem of sensor selection for state estimation of PCCP, and we demonstrate how the cost of sensors may also be considered in sensor selection for PCCP and generic nonlinear processes while maintaining high computational efficiency. First, a local sensitivity matrix is constructed based on the candidate sensors, which represents the sensitivities between the states and the candidate output measurements. Then, a new performance index is defined based on both the degree of observability and the economic cost of the candidate sensors. The designed performance index is positively related to the observability of the system and negatively correlated with the sensor cost. Moreover, the sensor selection problem is formulated as an optimization problem, and we propose a computationally efficient method to solve it such that sensors are placed optimally for the PCCP. Additionally, we consider the cases when sensor malfunctions may take place and address a resilient sensor selection problem for the PCCP. Our objective is to maximize the observability of the entire process considering the worst case scenarios given the remaining healthy sensors, while still meeting the budget constraint. Good state estimation results are obtained based on the recommended sets of sensors in both cases. 

\section{Problem formulation}

\subsection{Model description}
\label{Model}

\begin{figure}[!hbt]
	\centering
	\includegraphics[width=0.8\hsize]{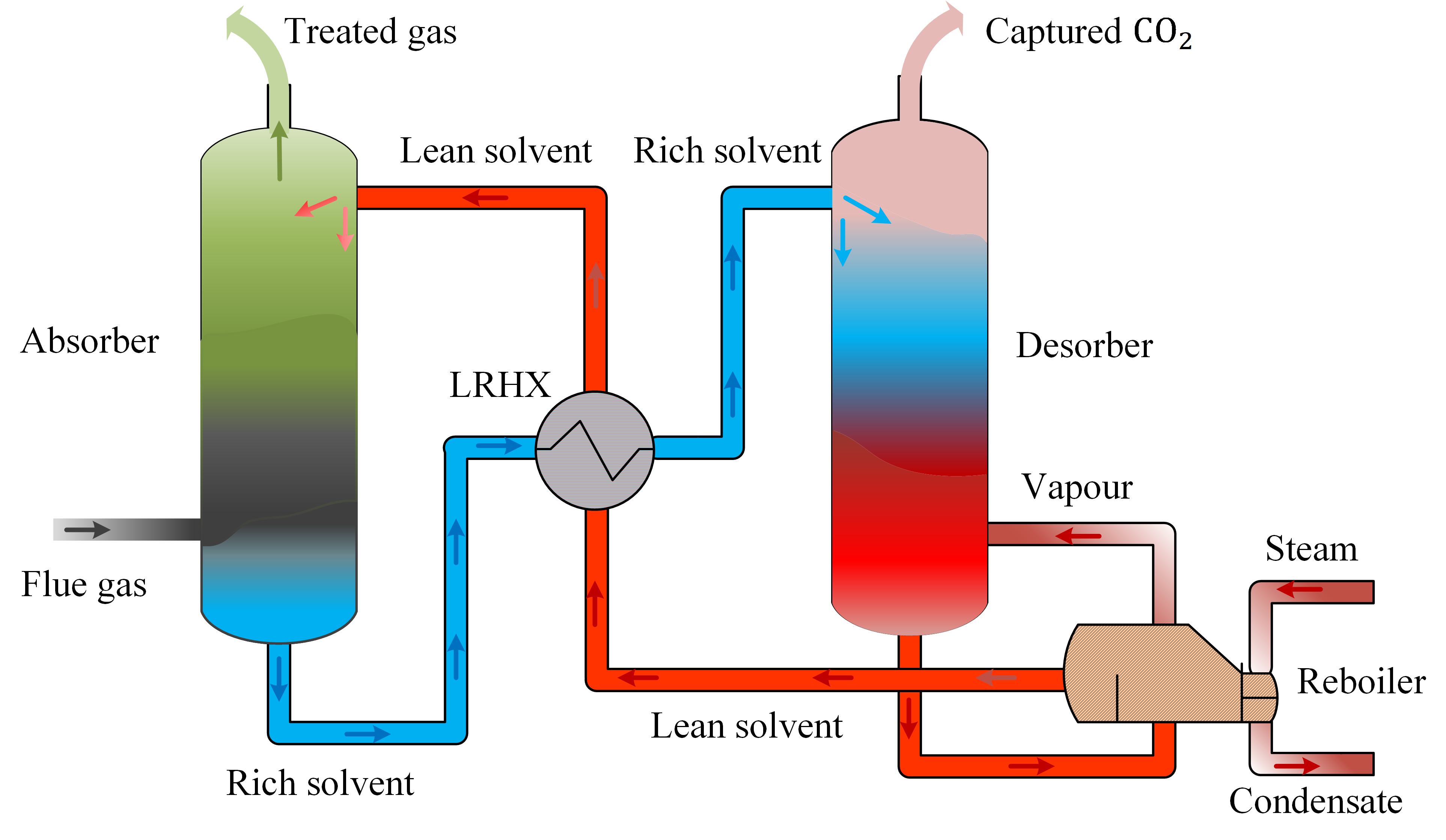}
	\caption{Flow diagram of an amine-based post-combustion CO$_2$ capture plant.}
	\label{PCCPlant_flowchart}
\end{figure}
The post-combustion CO$_2$ capture plant considered in this paper is depicted in Figure~\ref{PCCPlant_flowchart}. This diagram shows the four primary components of the PCCP, namely the absorption and desorption columns, lean-rich heat exchanger (LRHX), and reboiler. The flue gas (containing a substantial amount of carbon dioxide CO$_2$) enters the bottom of the absorber from the power plant and is contacted with the lean solvent -- solvent with low amounts of CO$_2$. This study uses 5M Monoethanolamine (MEA) as the solvent. The treated flue gas leaves the absorption column with a low concentration of CO$_2$. The rich solvent containing a high amount of CO$_2$ goes through the LRHX, exchanging heat with the lean solvent coming out of the reboiler. The rich solvent then moves into the top of the desorption column, where it is heated by contacting the reboiler's hot vapor. In the desorption column, the CO$_2$ is stripped from the rich solvent that is recycled back to the absorber. The CO$_2$ gas with a high concentration of CO$_2$ (90–99$\%$) is discharged from the desorption column. 

The details of the model are briefly described in the following. First, the models for the absorber and desorption column are similar, except for a few details like the direction of reactions, temperature, and reaction rate constants. The partial differential equations (\ref{lsy11_2.1a})--(\ref{lsy11_2.1d}) describe the dynamic model for the two columns, which are derived based on mass and energy balances \cite{Decardi-Nelson2018_Process,Harun2012_IJGGC}:
\begin{align}
	\label{lsy11_2.1a}
	\frac{\partial C_L(i)}{\partial t} &= \frac{F_{L}}{S_c}\frac{\partial C_L(i)}{\partial l} +(N(i)a^{I}),\; i=CO_2, MEA, H_2O, N_2,\\	
	\label{lsy11_2.1b}
	\frac{\partial C_G(i)}{\partial t} &= -\frac{F_{G}}{S_c}\frac{\partial C_G(i)}{\partial l} -(N(i)a^{I}), \\
	\label{lsy11_2.1c}
	\frac{\partial T_L}{\partial t} &= \frac{F_{L}}{S_c}\frac{\partial T_L}{\partial l} +\frac{(Q_La^I)}{\sum_{i=1}^{n}C_L(i)C_{p,i}}, \\
	\label{lsy11_2.1d}
	\frac{\partial T_G}{\partial t} &= -\frac{F_{G}}{S_c}\frac{\partial T_G}{\partial l} +\frac{(Q_Ga^I)}{\sum_{i=1}^{n}C_G(i)C_{p,i}},
\end{align}	
where the subscripts $L$ and $G$ denote liquid and gas phases, respectively. The definitions of other variables and parameters are shown in Table~\ref{lsy11_tab_variables}. In (\ref{lsy11_2.1a})--(\ref{lsy11_2.1d}), the dependent variables change with the time $t$ and axial position $l$ of the column. It is also assumed that each stage in the two columns is well mixed. 

\begin{table} 
	\centering
	\caption{Process variables and parameters of the absorption and desorption columns.}
	\label{lsy11_tab_variables}
	\renewcommand{\arraystretch}{1.2}
	\tabcolsep 33pt
	\begin{tabular}{lll}\hline
	    Notation & Definition & Unit   \\\hline
		$C(i)$   & molar concentrations of component $i$ & $\rm mol/m^3$ \\
		$S_c$    & cross-sectional area of the column & $\rm m^2$ \\
		$F$      & volumetric flow      & $\rm m^3/s$ \\
		$N(i)$   & mass transfer rate   & $\rm kmol/m^2s$ \\
		$T$      & temperature          & $\rm K$  \\
		$l$      & height of the column & $\rm m$  \\
		$C_p$    & heat capacity        & $\rm KJ/kmol$ \\
		$Q$      & heat transfer rate   & $\rm KJ/m^2s$ \\
		$a^{I}$  & interfacial area     & $\rm m^2/m^3$ \\
		$V$      & volume               & $\rm m^3$     \\
		$\rho$   & average molar density& $\rm kmol/m^3$\\
		$H$      & enthalpy             & $\rm KJ$      \\
		$f$      & flow rate            & $\rm mol/s$   \\\hline
\end{tabular}\end{table}

Another important unit of the PCCP is the lean-rich heat exchanger. The heat exchanger transfers heat from the hot lean solvent to the cold solvent from the absorber making the process heat efficient. It is assumed that the mass inside the heat exchanger remains constant. Therefore, the energy balance equations in (\ref{lsy11_2.1e})--(\ref{lsy11_2.1f}) are used to represent the dynamics of the heat exchanger,
\begin{align}
\label{lsy11_2.1e}
	\frac{dT_{tube}}{dt} &= \frac{\dot{V}_{tube}}{V_{tube}}(T_{tube,in}-T_{tube,out}) + \dot{Q}\frac{1}{C_{p_{tube}}\rho_{tube}V_{tube}},\\
\label{lsy11_2.1f}
	\frac{dT_{shell}}{dt} &= \frac{\dot{V}_{shell}}{V_{shell}}(T_{shell,in}-T_{shell,out}) + \dot{Q}\frac{1}{C_{p_{shell}}\rho_{shell}V_{shell}},
\end{align}
where $\dot{V}(\rm m^3/s)$ and $\dot{Q}(\rm kJ/s)$ represent the volumetric flow and heat transfer rate, respectively, the subscripts $tube$, $shell$, $in$ and $out$ denote the tube-side, shell-side, inlets and outlets of the heat exchanger, respectively. Table~\ref{lsy11_tab_variables} shows the definitions of the other variables and parameters.

The reboiler provides heat to the whole plant. The rich solvent is heated in this unit to break the chemical bonds between CO$_2$ and MEA. During this procedure, water, carbon dioxide, and MEA are converted to vapor and moved to the bottom of the desorption unit. In this study, we assumed that the liquid level and the reboiler pressure does not change. Thus, there is no vapour or liquid holdup. The energy balance equation is listed as follows:
\begin{align}
\label{lsy11_2.1g}
	\rho C_pV\frac{dT_{reb}}{dt} = f_{in}H_{in}-f_{V}H_{V,out}-f_{L}H_{L,out}+Q_{reb},
\end{align}
where $T_{reb}(\rm K)$ represents the temperature of reboiler, the subscripts $in$, $out$, $V$, and $L$ denote inlet, outlet, vapour and liquid, respectively, $Q_{reb}(\rm KJ/s)$ is the heat input. The definitions of the other variables and parameters can be found in Table~\ref{lsy11_tab_variables}. 

Equations (\ref{lsy11_2.1a})--(\ref{lsy11_2.1g}) are the modeling equations of the PCCP. For the model development, physical property calculations of gas and liquid phases are necessary for the heat and mass transfer process, which are estimated from seven nonlinear algebraic correlations. These specific calculations not included in this work and can be found in \cite{Decardi-Nelson2018_Process,Harun2012_IJGGC}.

\subsection{Model discretization}

According to the model of the PCCP, the two columns are formulated as partial differential equations, while the heat exchanger and reboiler are formulated as ordinary differential equations. Additionally, some model parameters are calculated using algebraic equations. Since the variables in the model of the two columns show temporal and spatial distributions, the partial differential equations are discretized into ordinary differential equations using the method in \cite{Decardi-Nelson2018_Process}. Specifically, the derivatives with respect to the length of the column are discretized into five stages. Therefore, the model described in Section \ref{Model} can be expressed as a system of differential algebraic equations:
\begin{align}
\label{lsy11_2.2a}
	\dot{\bfx}(t)&=\bfF(\bfx(t),\bfa(t),\bfu(t)),\\
\label{lsy11_2.2b}
	\bf0&=\bfH(\bfx(t),\bfa(t),\bfu(t)),
\end{align}
where $\bfx\in\R^{103}$ is the differential state vector, and the definitions of state variables are shown in Table~\ref{lsy11_taba_sensor}, $\bfa\in\R^7$ is the algebraic state vector, where $a_1$--$a_4$ are the mole fractions of the four compositions in liquid phase in the reboiler, $a_5$ is the reboiler vapour fraction, $a_6$ is the CO$_2$ concentration of the liquid entering the absorption column, $a_7$ is the flowrate of gas entering the desorption column, $\bfu=[F_L, Q_{reb}, F_G]\in\R^3$ denotes the input vector: solvent flow rate in L/s, reboiler heat in KJ/s, and flue gas flow rate in m$^3$/s. 

\begin{table}[t] 
	\centering
	\caption{Definition of the state variables of each unit.}
	\label{lsy11_taba_sensor}
	\renewcommand{\arraystretch}{1.2}
	\tabcolsep 9pt
	\begin{tabular}{cc|cc|cc|cc}\hline
		\multicolumn{2}{c|}{Absorber} & \multicolumn{2}{c|}{Desorber} & 
		\multicolumn{2}{c|}{Heat exchanger} &\multicolumn{2}{c}{Reboiler}\\\hline
		States      & Definition & States & Definition & States & Definition & States & Definition\\\hline
		$x_{1-5}$   & $C_{L}^j$(N$_2$)  & $x_{51-55}$ & $C_{L}^j$(N$_2$)  & $x_{101}$ & $T_1$ & $x_{103}$ & $T$ \\
		$x_{6-10}$  & $C_{L}^j$(CO$_2$) & $x_{56-60}$ & $C_{L}^j$(CO$_2$) & $x_{102}$ & $T_2$ &  &  \\
		$x_{11-15}$ & $C_{L}^j$(MEA)    & $x_{61-65}$ & $C_{L}^j$(MEA)    &&&&\\
		$x_{16-20}$ & $C_{L}^j$(H$_2$O) & $x_{66-70}$ & $C_{L}^j$(H$_2$O) &&&&\\
		$x_{21-25}$ & $T_{L}^j$         & $x_{71-75}$ & $T_{L}^j$         &&&&\\
		$x_{26-30}$ & $C_{G}^j$(N$_2$)  & $x_{76-80}$ & $C_{G}^j$(N$_2$)  &&&&\\
		$x_{31-35}$ & $C_{G}^j$(CO$_2$) & $x_{81-85}$ & $C_{G}^j$(CO$_2$) &&&&\\
		$x_{36-40}$ & $C_{G}^j$(MEA)    & $x_{86-90}$ & $C_{G}^j$(MEA)    &&&&\\
		$x_{41-45}$ & $C_{G}^j$(H$_2$O) & $x_{91-95}$ & $C_{G}^j$(H$_2$O) &&&&\\
		$x_{46-50}$ & $T_{G}^j$         & $x_{96-100}$& $T_{G}^j$         &&&&\\\hline
	\end{tabular}
\end{table}

\section{Sensor selection for PCCP}

\subsection{Available sensors and costs}
\label{Sensor_cost}

The first step is to determine the types of sensors available for the PCCP. Typically, there may be concentration, flow, pressure and temperature sensors in the two columns to monitor the variables inside, as well as in the heat exchanger and reboiler. There may be some other sensors for equipment safety purpose. In this work, we only consider sensor selection for process state estimation and monitoring purpose. The system states are the temperatures of these units and the concentrations of CO$_2$, MEA, H$_2$O, and N$_2$. Among these four substances, commercially available CO$_2$ concentration sensors make them suitable for this purpose. Therefore, we consider CO$_2$ concentration and temperature sensors as available sensors for this study. 

\begin{figure}[!hbt]
	\centering
	\includegraphics[width=0.9\hsize]{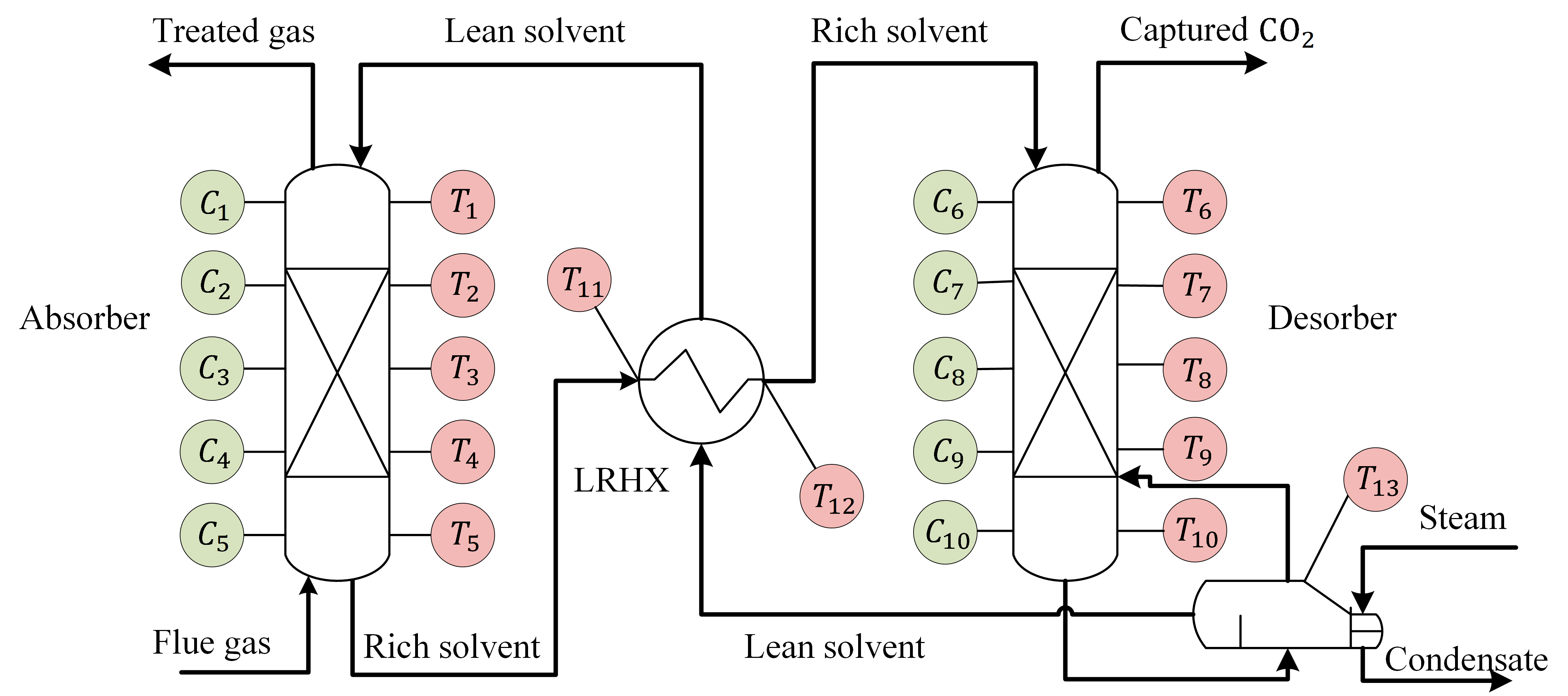}
	\caption{Available concentration and temperature sensors installed on PCCP.}
	\label{lsy11_Sensorplacement}
\end{figure}

Figure~\ref{lsy11_Sensorplacement} shows candidate sensors and their locations for the PCCP. Five equidistant points on the height of the absorption and desorption columns are selected for the placement of temperature and concentration sensors, as shown in the figure. Concentration sensors for absorber are denoted as $C_1$--$C_5$, and temperature sensors are denoted as $T_1$--$T_5$. Similarly, $C_6$--$C_{10}$ and $T_6$--$T_{10}$ are concentration and temperature sensors for desorption column, respectively. $T_{11}$ and $T_{12}$ denote the temperature sensors for the heat exchanger, and $T_{13}$ is the temperature sensor for the reboiler. Therefore, there are in total 23 candidate sensors, which are expressed as a set $\mathcal{S}^{(23)}=\{C_1, \dots, C_5, T_1, \ldots, T_5, C_6, \ldots, C_{10}, T_6, \ldots, T_{13}\}$. The measurable states in each unit of the PCCP are listed in Table~\ref{lsy11_taba_sensors}. The cost of sensors depends mainly on their type and number. It is considered that the cost of one temperature sensor is about 1,000 USD, while the cost of one concentration sensor is more expensive which is about 20,000 USD.

\begin{table}[t] 
	\centering
	\caption{Available sensor locations in each unit of the PCCP.}
	\label{lsy11_taba_sensors}
	\renewcommand{\arraystretch}{1.1}
	\tabcolsep 4pt
	\begin{tabular}{l|c|c}\hline
		Unit      & Sensors & No.  \\\hline
		Absorber  & $C_{G}^1$(CO$_2$), $C_{G}^2$(CO$_2$), $C_{G}^3$(CO$_2$), $C_{G}^4$(CO$_2$), 
		            $C_{G}^5$(CO$_2$), $T_{G}^1$, $T_{G}^2$, $T_{G}^3$, $T_{G}^4$, $T_{G}^5$ & 10 \\
		Desorber  & $C_{G}^1$(CO$_2$), $C_{G}^2$(CO$_2$), $C_{G}^3$(CO$_2$), $C_{G}^4$(CO$_2$), 
		            $C_{G}^5$(CO$_2$), $T_{G}^1$, $T_{G}^2$, $T_{G}^3$, $T_{G}^4$, $T_{G}^5$ & 10 \\
   Heat exchanger & $T_1$, $T_2$ & 2 \\
	    Reboiler  & $T$ & 1 \\\hline
	\end{tabular}
\end{table}

\subsection{Sensor selection problem}

It is assumed that we want to select a subset of sensors from the 23 candidate sensors so that the entire system state is observable and the cost of the sensors is within a given budget constraint. Our objective is to improve the state estimation performance and minimize the redundancy among the selected sensors when solving the cost-constrained sensor selection problem. 

When a sensor in the set of candidate sensors $\mathcal{S}^{(m)}$ ($m=23$) is placed, it provides a measurement $y_i(k)$. The following equation represents the measurement equation provided by all the candidate sensors:
\begin{equation}
\label{lsy11_2.2c}
	\bfy(k)=\bfC\bfx(k)+\bfv(k),
\end{equation}
where $\bfy(k)\in\R^{m}$ is the measurement vector, $\bfv(k)\in\R^{m}$ is the measurement noise and $\bfC\in\R^{m \times n}$ denotes the measurement matrix. The cost of each sensor $i\in\mathcal{S}^{(m)}$ is denoted by $g_i\in\R_{\geqslant0}$, which represents the purchase and installation costs associated with the type of sensor used. Define the cost vector as $\bfg=[g_1, \ldots, g_m]^{\T}$. Assume that the total cost spent on sensors cannot be more than a given sensor budget $G\in{R}_{\geqslant0}$. Let $\bfz\in\{0, 1\}^{m}$ be a binary decision vector to denote which sensors are selected; that is, $z_i=1$ if $i\in\mathcal{S}^{(m)}$ is selected and otherwise $z_i=0$. 

The sensor selection problem is to find the indicator vector $\bfz$ of the sensors and can be formulated as the following optimization problem:
\begin{align}
	& \max_{\bfz\in\{0,1\}^m} { J(\bfz) } \label{eq:obj},\\
	& {\rm s.t.} \quad  \bfg^{\T} \bfz \leq G \label{eq:cost},
\end{align}
where $J(\bfz)$ is a performance index, $\bfg^{\T}\bfz$, denoted as $c_{actual}$, is the actual cost for installing the selected sensors. 

In sensor selection, the performance index $J(\bfz)$ is typically designed to reflect the degree of observability of the system provided by the selected sensors or the state estimation performance that can be achieved from the selected sensors. For linear system, a popular design of the performance index is the trace of the steady-state error covariance matrix of the corresponding Kalman filter \cite{Zhang2017_Auto}, which directly reflects the estimation performance. However, for nonlinear systems, sensor selection is more challenging. It is generally preferable to use a performance index that reflects the degree of observability instead of the estimation performance since the estimation performance of nonlinear estimators/filters is generally not computationally easy to obtain. Moreover, for nonlinear systems, the optimization problem in (\ref{eq:obj})--(\ref{eq:cost}) is a nonlinear integer optimization, which is typically computationally very expensive to solve. In the next section, we will propose a performance index design that reflects the degree of observability provided by the selected sensors, for which a computationally efficient solving approach can be developed.

\section{Proposed sensor selection method for PCCP}
\label{Section 4}

In this section, we first introduce how sensitivity analysis can be used to assess the degree of observability of a system given a set of sensors. Then, we formulate an objective function for the sensor selection for the PCCP. The formulated constrained optimization problem has been approximated and can be efficiently solved.

\subsection{Calculation of degree of observability}

Sensitivity analysis is a useful tool for quantifying the dependence of the model outputs on the model parameters, which can be applied for various purposes, such as model identification and model reduction \cite{Guo2021_JPC}. For the nonlinear systems, the local observability of the state can be measured using the sensitivity matrix of the outputs to the initial state $\bfx(0)$ \cite{Liu2021_IECR,LiuSY2022_ChERD}. In \cite{Liu2021_IECR,LiuSY2022_ChERD}, this approach has been applied to systems described by ordinary differential equations (ODEs). However, the first-principles model of the PCCP consists of differential-algebraic equations (DAEs). Therefore, we extend this method to the context of nonlinear DAE processes to assess the degree of observability of the PCCP in \eqref{lsy11_2.2a} and \eqref{lsy11_2.2b}.

It is assumed that the initial state $\bfx(0)$ for the DAE system described by (\ref{lsy11_2.2a}) and (\ref{lsy11_2.2b}) is known or can be estimated. The model of the PCCP is discretized and expressed as the corresponding system: 
\begin{align}
	\label{lsy11_4a}
	\bfx(k+1)&=\bff(\bfx(k),\bfa(k),\bfu(k)),\\
	\label{lsy11_4b}
	\bfa(k)&=\bfphi(\bfx(k),\bfu(k)).
\end{align}
The algebraic state vector $\bfa(k)$ is only dependent on $x_{103}(k)$ and can be expressed as (\ref{lsy11_4b}). With the known initial condition, $\bfa(0)=\bfphi(\bfx(0),\bfu(0))$ can be calculated based on (\ref{lsy11_4b}). By substituting $\bfa(0)$ into (\ref{lsy11_4a}), we can obtain $\bfx(1)=\bff(\bfx(0),\bfa(0),\bfu(0))$. By performing the above steps recursively, $\bfa(k)$ and $\bfx(k+1)$ can be calculated by (\ref{lsy11_4a}) and (\ref{lsy11_4b}). The sensitivity matrix can be constructed from the initial time instant to an arbitrary sampling instant $k>0$ as:
\begin{align}
	\label{lsy11_4.1c}
	\bfS(k,0)=\vectf{\bfS_{y,x(0)}(0)}{\bfS_{y,x(0)}(1)}{\vdots}{\bfS_{y,x(0)}(k)}\in\R^{(k+1)m\times n},
\end{align}
where $\bfS_{y,x(0)}(k)\in\R^{m\times n}$ at time $k$ is calculated by 
\begin{align}
\label{lsy11_4.1d}
	\bfS_{y,x(0)}(k)=\frac{\partial \bfy}{\partial \bfx}(k)&=\bfC\frac{\partial \bff}{\partial \bfx}(k-1)\frac{\partial \bff}{\partial \bfx}(k-2)\cdots\frac{\partial \bff}{\partial \bfx}(0)\non\\
	&=\bfC\frac{\partial \bff}{\partial \bfx}\big|_{
	\begin{smallmatrix}
		\bfx=\bfx(k-1) \\{\bfa=\bfa(k-1)}
	\end{smallmatrix}}\frac{\partial \bff}{\partial \bfx}\big|_{
    \begin{smallmatrix}
        \bfx=\bfx(k-2) \\{\bfa=\bfa(k-2)}
    \end{smallmatrix}}\cdots\frac{\partial \bff}{\partial \bfx}\big|_{
    \begin{smallmatrix}
	    \bfx=\bfx(0) \\{\bfa=\bfa(0)}
    \end{smallmatrix}},
\end{align}
with the initial value $\frac{\partial \bff}{\partial \bfx}(0)=\bfI$. The above $\bfS_{y,x(0)}(k)$ provides the sensitivity information based on the measurements obtained at time $k$. Specifically, the sensitivity of different outputs to each state variable is represented by the elements in each column of the time-dependent matrix $\bfS_{y,x(0)}(k)$. 

Based on the local sensitivity matrix $\bfS(k,0)$, we can quantify the amount of information provided by the measurements towards estimating the initial state through orthogonalization, as proposed in our previous work \cite{LiuSY2022_ADCONIP}. In this work, we refer to the amount of information as the degree of observability provided by the measurements or the corresponding sensor set. Specifically, for the sensitivity matrix $\bfS(k,0)$, to calculate the corresponding degree of observability, we first calculate the norm of each of the column vectors and denote the largest one as $N_1$. The remaining $n-1$ columns are then projected onto the column vector with the largest norm. The residuals of the $n-1$ column vectors are all orthogonal to the largest vector. From these $n-1$ residual vectors, the above procedure is repeated. That is, we calculate the norms of these residual vectors and find the residual vector with the largest norm. Its norm is denoted as $N_2$. The remaining $n-2$ vectors are then projected onto the vector corresponding to $N_2$ to find the remaining residual vectors and the above procedure continues to find $N_i$, $i=3,\dots, n$. The degree of observability is defined as the summation of $N_i$, $i=1,\dots, n$ as follows:
\begin{align}
\label{lsy11_4.1e}
	\lambda=\left\{
	\begin{aligned}
		&\sum_{i=1}^{n}N_i, &  & {\rm rank}(\bfS(k,0)) = n, \\
		&0, &  & {\rm rank}(\bfS(k,0)) < n.
	\end{aligned}\right.
\end{align}
If rank($\bfS(k,0))=n$, then the system is observable, and the degree of observability is given by $\lambda=\sum_{i=1}^{n}N_i$. If rank($\bfS(k,0))\textless n$, then the system is unobservable, and $\lambda$ is set to zero. This implies that we should only evaluate degree of observability of observable systems. After defining $\lambda$, we can design the objective function based on it in the next subsection.

\subsection{Objective function based on degree of observability }

When selecting sensors for the PCCP, two factors are taken into account. First, the selected sensors should provide a high degree of observability for the system. Second, the cost of the selected sensors must fit within the total budget allocated for sensors.

To incorporate the degree of observability $\lambda(\bfz)$ as the objective function in place of the performance index $J(\bfz)$, the sensor selection optimization problem in (\ref{eq:obj})--(\ref{eq:cost}) can be more specifically written as:
\begin{align}
\label{lsy11_4.2f}
	& \max_{\bfz}\lambda(\bfz),\\
	\label{lsy11_4.2g}
    {\rm s.t.} & \ \bfg^{\T} \bfz \leq G,\\
\label{lsy11_4.2g1}
	& \  \lambda(\bfz) > 0.
\end{align}
Equations (\ref{lsy11_4.2f})--(\ref{lsy11_4.2g1}) describe the cost-constrained sensor selection problem, which aims to maximize the degree of observability measured by (\ref{lsy11_4.1e}), subject to the budget constraint and observability requirement. This is a nonlinear integer programming problem. Since there is a large number of candidate sensors that may be deployed for the PCCP, the number of integer decision variables is also large, which makes the associated constrained optimization problem in the form of (\ref{lsy11_4.2f})--(\ref{lsy11_4.2g1}) computationally expensive. 

Based on the above consideration, instead of solving the optimization directly, we propose a heuristic method to address the cost-constrained sensor selection problem in a computationally efficient manner. We explicitly incorporate the cost of sensors into the objective function and define a new objective function as follows:
\begin{align}
\label{lsy11_4.2h}
	J_1(\bm z) = \frac{ \lambda(\bm z) } {\bm g^{\rm T} \bm z} = {\rm CPI}(\bm z),
\end{align}
where ${\rm CPI}(\bm z)$ is defined as the cost performance index.  When maximizing $J_1(\bm z)$, the optimization aims to maximize the degree of observability while reducing the sensor cost. Using the new performance index, the original constrained integer optimization shown in \eqref{eq:obj}--\eqref{eq:cost} or \eqref{lsy11_4.2f}--\eqref{lsy11_4.2g1} is approximately transformed into the following problem:
\begin{align}
\label{lsy11_4.2i}
	& \max_{\bm z} { \frac{ \lambda(\bm z) } {\bm g^{\rm T} \bm z} },\\
\label{lsy11_4.2i1}
	& {\rm s.t.}\  \lambda(\bfz) > 0.
\end{align}
The above optimization seeks a sensor set that maximizes the degree of observability while minimizing the sensor cost. 

Note that the solution to the optimization problem in (\ref{lsy11_4.2i})--(\ref{lsy11_4.2i1}) may not be equivalent to the solution of the original optimization problem in \eqref{eq:obj}--\eqref{eq:cost} or \eqref{lsy11_4.2f}--\eqref{lsy11_4.2g1}. To reduce the gap between the original and the approximated optimization problems, a practical approach is to introduce a tuning parameter to the cost performance index (CPI) as follows:
\begin{align}
\label{lsy11_4.2j}
	J_2(\bfz) = \frac{ \lambda(\bfz) } {(\bfg^{\rm T} \bfz)^{\alpha}},
\end{align}
where $0<\alpha<\infty$. By obtaining solutions with different $\alpha$ values using the algorithm that will be discussed in the next subsection and selecting the solution that provides the best degree of observability, it is possible to obtain a near-optimal solution.

\begin{remark}
	Based on the definition of the objective function $J_2$ in Equation (\ref{lsy11_4.2j}), it can be found that when $\alpha=1$, $J_2$ is equal to $J_1$ defined in (\ref{lsy11_4.2h}). When $\alpha$ is 0, $J_2$ is equivalent to the degree of observability $\lambda(\bfz)$. In this case, the optimization problem seeks to maximize the degree of observability without considering the cost of sensors.
\end{remark}

\section{Solving the optimization problem}
\label{five}

In this section, we will discuss the method proposed to efficiently solve (\ref{lsy11_4.2i})--(\ref{lsy11_4.2i1}) and obtain the (near-)optimal set of sensors for the original optimization problem that satisfies the cost constraint. 

\subsection{Solving the sensor selection problem}

Before introducing the proposed method, some assumptions and initial conditions are required. Initially, all the candidate sensors in $\mathcal{S}^{(m)}$ are considered, and it is assumed that the system is observable when all candidate sensors are included. Under the initial conditions, the cost of sensors exceeds the budget, and it is also assumed that a subset exists that satisfies the cost budget and ensures observability. 

The subset of sensors is determined by removing sensors from the
set of candidate sensors one by one based on the CPI$(\bfz)$. Specifically, the sensor that contributes the least to the CPI$(\bfz)$ is identified and removed from the set of candidate sensors. To identify the sensor that contributes the least to the CPI$(\bfz)$, we first find all the $m$ subsets that contain $m-1$ sensors. For each of these $m-1$ sensor subset, we calculate the corresponding CPI$(\bfz)$ value. Among the $m$ CPI$(\bfz)$ values obtained, we identify the subset that gives the highest CPI$(\bfz)$ value. The sensor that is not included in the subset is the one that contributes the least to the CPI$(\bfz)$ of the sensor set $\mathcal{S}^{(m)}$. The sensor is then removed from the candidate sensor set $\mathcal{S}^{(m)}$, and we obtain the candidate sensor set $\mathcal{S}^{(m-1)}$, which contains $m-1$ sensors. If the sensor set $\mathcal{S}^{(m-1)}$ satisfies the sensor budget constraint, it is the solution. Otherwise, we carry out the above procedure to remove more sensors one by one until the sensor budget constraint is satisfied.

\begin{algorithm} \label{minimum_set}
	\caption{Remove sensor one-by-one by considering the CPI}
	\KwIn{Given the set containing all sensors: $\mathcal{S}^{(m)}=\{1,2,\ldots,m\}$, the cost threshold $G$, and the cost vector $\bfg$}
	
	$\mathcal{S}_{opt} \Leftarrow \mathcal{S}^{(m)}$ \\
	\While{$\bfg^{\T}\bfz>G$}{
		\For{$i\in\mathcal{S}_{opt}$}{
			$\mathcal{S}_i=\mathcal{S}_{opt}\backslash \{i\}$\\
			Check the observability and calculate CPI$(\bfz)$}
		\If{{\rm all CPI}$(\bfz)=0$}{
			Terminate the procedure.}
		\Else{
			Choose $i^*=\arg\max_i$CPI$(\bfz)$.\\
			$\mathcal{S}_{opt}=\mathcal{S}_{i^*}$ }}
	\KwOut{The selected sensor set $\mathcal{S}_{opt}$}
\end{algorithm}
Algorithm~\ref{minimum_set} summarizes the sensor selection procedure discussed above. The algorithm can help exclude candidate sensors from the set of selected sensors that have a low price-performance ratio. Specifically, the proposed algorithm tend to remove sensors that are very expensive yet do not contribute significantly to the increase in system observability. With Algorithm~\ref{minimum_set}, the observability test needs to be conducted for entire removing process. When we find that the system is no longer observable after removing a sensor, we set the degree of observability $\lambda(\bfz)$ to zero so that this sensor will remain selected. The CPI$(\bfz)$ is zero in the case of $\lambda(\bfz)$ is zero. This implies that we should only evaluate the CPI$(\bfz)$ of observable systems.

\begin{remark}
	It is widely recognized that computational complexity is a major concern when it comes to sensor selection. The proposed algorithm addresses this issue by reducing the number of sensor combinations that need to be considered. Specifically, when selecting $r$ sensors from a set of $m$ sensors, the number of possible sensor combinations that need to be evaluated is $m+(m-1)+\dots+(r+1)+r=\frac{1}{2}(m-r+1)(m+r)$. This results in a computational complexity of $\mathcal{O}(m^2)$, which is more efficient than solving integer programming. In comparison, the computational complexity of mixed-integer linear programming is $\mathcal{O}(2^m)$. Therefore, the proposed algorithm offers significant computational advantages for solving the sensor selection problem.  
\end{remark}

\begin{remark}
	In the proposed method, the subset of sensors is obtained by iteratively decreasing the number of sensors in the set of candidate sensors until a set of sensors that satisfies the cost constraint is obtained. Note that this method is not guaranteed to provide the globally optimal solution, but it can provide a near-optimal solution.
\end{remark}

\begin{figure}[!hbt]
	\centering
	\includegraphics[width=0.8\hsize]{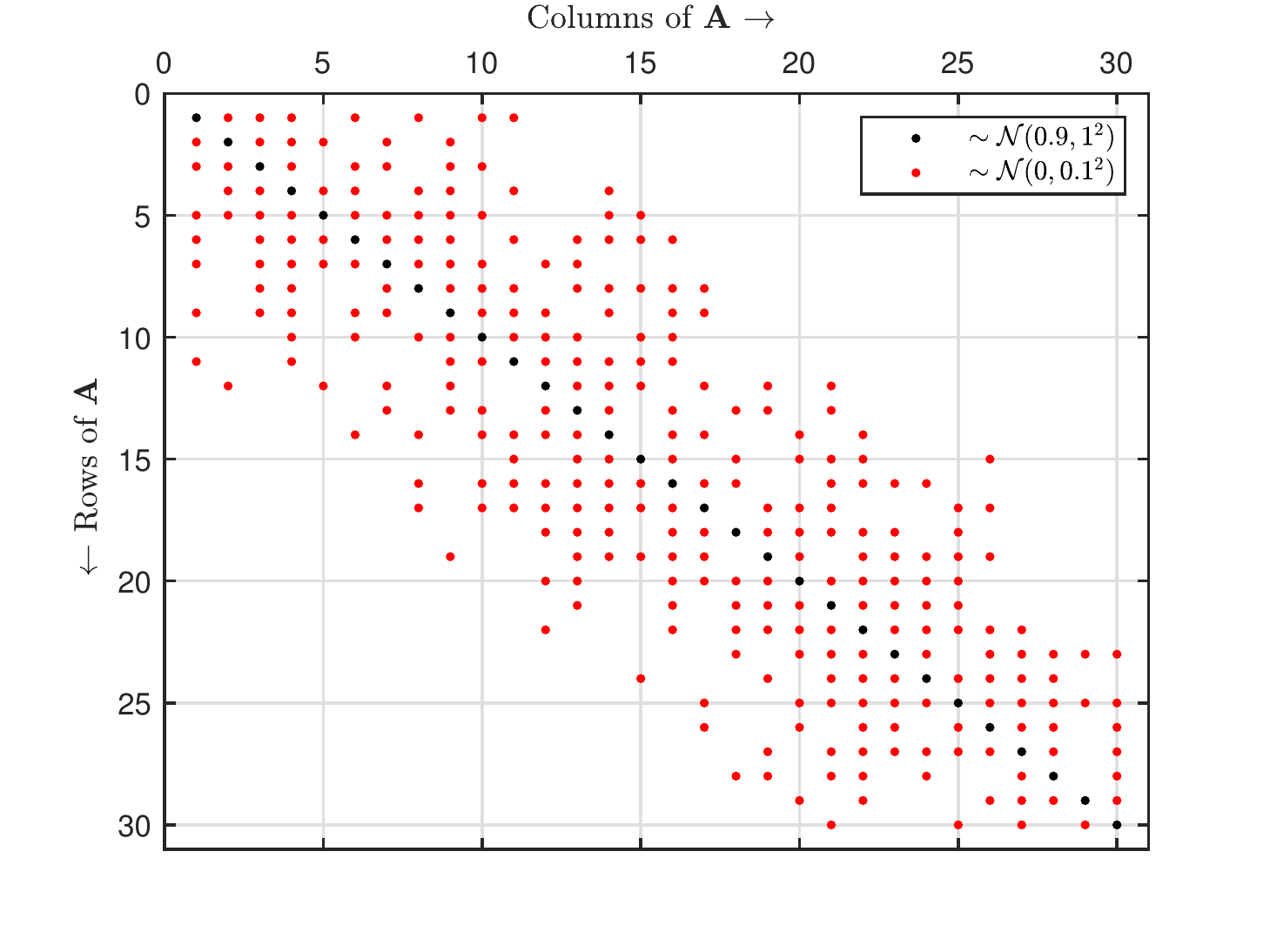}
	\caption{Graphical representation of the non-zero elements of system matrix $\bfA$ of Example 1.}
	\label{lsy11_fig_A}
\end{figure}

\paragraph{\bf A numerical Example.} In order to test the applicability of our proposed generalized cost performance index, we consider a randomly generated linear system with 30 states. The system matrix $\bfA\in\R^{30\times 30}$ is described by Figure~\ref{lsy11_fig_A}, where each dot (in either red or black) represents a non-zero value at the corresponding position. The values of the elements of $\bfA$ matrix characterized by red dots are randomly generated following a Gaussian distribution $\mathcal{N}(0,0.1^2)$. The values of the system matrix elements associated with the black dots are generated following a Gaussian distribution $\mathcal{N}(0.9,1^2)$. Initially, the output measurement matrix is $\bfC=\bfI_{30}$, assuming that all the states are measurable in this example. The cost of each sensor is generated following a Gaussian distribution $g_i \sim \mathcal{N}(10,2^2)$ and is then bounded within the range [6.0, 13.8].

\begin{figure}[!hbt]
	\centering
	\includegraphics[width=\hsize]{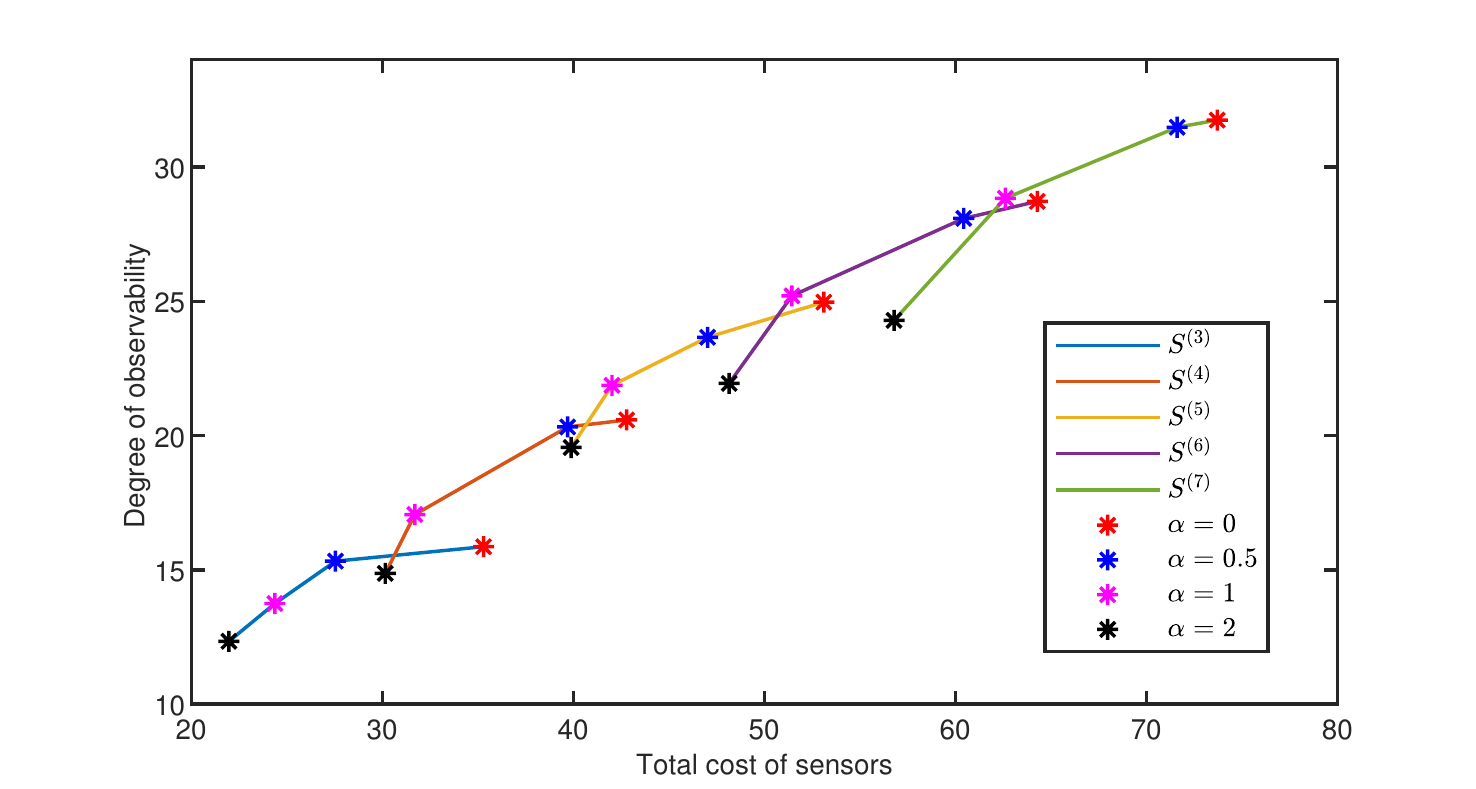}
	\caption{Relation between degree of observability and cost of sensors under different $\alpha$ and numbers of sensors.}
	\label{lsy11_fig_degVScost}
\end{figure}

Then, we apply the proposed method to this numerical example. Figure~\ref{lsy11_fig_degVScost} depicts the relationship among the degree of observability, total cost of sensors, the value of $\alpha$, and the number of sensors being considered. The figure includes five lines, each of these lines represents the selection results (degree of observability and cost of selected sensors) with a fixed number of selected sensors but with different $\alpha$ values. For example, the blue line corresponds to $\mathcal{S}^{(3)}$, indicating that there are three sensors in the set. The four stars on each line represent different values of $\alpha$. 

As shown in Figure~\ref{lsy11_fig_degVScost}, the degree of observability increases with the number of sensors that can be used, but it decreases as the value of $\alpha$ increases. Figure~\ref{lsy11_fig_degVScost} also demonstrates that with smaller $\alpha$ values, the proposed method tends to place less emphasis on the cost of the sensors. When the cost index $\alpha$ gradually increases, which indicates that more emphasis is placed on the sensor cost, more cost-effective sensors tend to be chosen. As a result, the expenditure on sensors is reduced, which however leads to reduction in the degree of observability. These results show the proposed method can strike a balance between the degree of observability and the cost of sensors by adjusting the parameter $\alpha$ and the number of sensors. Specifically, $\alpha$ can be tuned to meet specific requirements for the cost and/or desired observability degree. 

\subsection{Robust sensor selection}

In general, Algorithm 1 provides a (near-)optimal solution for a given budget. However, certain regions in the PCCP operate under harsh environments, such as high temperature, steam and corrosive environments, which may cause the sensors assigned to such locations to fail. Sensor malfunction may damage the process equipment and cause a range of problems associated with the plant operations. One way to avoid sensor malfunctions is to place them in safer locations without direct exposure to extreme conditions, but this can result in a reduction in measurement accuracy. Therefore, it is worth investigating robust sensor selection when considering sensor malfunction, if there is an extra budget in addition to the budget for normal sensor selection, as discussed in the previous sections.

In this subsection, we further consider the scenario where sensors may malfunction to study fault tolerance (resilience) in sensor selection. Specially, we aim to form a resilient sensor selection such that the sensor network can provide good estimation performance even when some of the sensors fail. In this paper, we consider that up to $R\in\mathbb Z_{\geq0}$ sensors may fail. We introduce the indicator vector $\bfmu\in\{0, 1\}^{q}$ to indicate whether a sensor is malfunctioned. That is, $\mu_j=1$ if the $j$th sensor is malfunctioned; otherwise $\mu_j=0$ ($q$ is the number of sensors in $\mathcal{S}_{opt}$, $j=1,2,\dots,q$). Another indicator vector $\bfe\in\{0, 1\}^{m}$ indicates which additional sensors are selected to enhance the robustness of the sensor selection. That is, $e_i=1$ if the $i$th sensor is selected extra; otherwise $\mu_i=0$.

The resilient sensor selection problem is formulated as follows:
\begin{align}
\label{lsy11_5.2a}
	& \max_{\bfz+\bfe}\min_{\bfmu}\lambda(\bfz+\bfe-\bfmu),\\
\label{lsy11_5.2bb}
	& {\rm s.t.} \quad  \bfg^{\T} \bfz \leq G, \\
\label{lsy11_5.2b}
	& \quad \quad \ \bfg^{\T} \bfe \leq G_e, \\
\label{lsy11_5.2b1}
	& \quad \quad \ \bfmu^{\T} \bfmu \leq R,\\
\label{lsy11_5.2b2}
	& \quad \quad \ \lambda(\bfz+\bfe-\bfmu) > 0,
\end{align} 
where $G_e$ is the extra budget constraint. It is considered that the total budget of resilient sensor selection can be divided into two parts: $G$ for sensor selection without considering robustness (as in Algorithm 1) and $G_e$ for improving the resilience of the sensor selection. Equations (\ref{lsy11_5.2a})--(\ref{lsy11_5.2b2}) describe the resilient sensor selection problem, which is a max-min optimization problem. It is assumed that a solution indeed exists to the above optimization problem. The solution is obtained by maximizing the minimum value of the degree of observability subject to sensor budget as in (\ref{lsy11_5.2bb}) and (\ref{lsy11_5.2b}), the limit on the number of malfunction sensors as in (\ref{lsy11_5.2b1}), and the observability requirement as in (\ref{lsy11_5.2b2}). Max-min problems are commonly used to define robust solutions and most of these problems turns out to be NP-hard \cite{Aissi2009_EJOR}. For the PCCP, there are a very large number of possible results that need to be considered, which is computationally expensive. Based on the above consideration, we propose a heuristic method to address the max-min problem in (\ref{lsy11_5.2a})--(\ref{lsy11_5.2b2}) in a computationally efficient manner.

The near-optimal solution to the resilient sensor selection problem is based on the solution to the normal sensor selection problem with the budget constraint $G$, denoted as $\mathcal{S}{opt}$, which can be obtained from Algorithm 1. The proposed solution is to add sensors one by one to $\mathcal{S}_{opt}$ based on the degree of observability. Specifically, the sensor that provides the maximum worst-case observability is identified and added to the set $\mathcal{S}_{opt}$. To identify the sensor to be added, we first select an arbitrary sensor $i$ from $\mathcal{S}^{(m)}$ to $\mathcal{S}_{opt}$. We define the updated set as $\mathcal{S}_i=\mathcal{S}_{opt}\cup \{i\}$, which contains $q+1$ sensors. We then remove one sensor $j\in\mathcal{S}_{i}$ and calculate the corresponding $\lambda_{ij}$ value based on $\mathcal{S}_{i} \backslash \{j\}$ for all $q+1$ possible cases. Among the $q+1$ $\lambda_{ij}$ values, we choose the minimum one: $\lambda_{i}=\min_j\lambda_{ij}$. Since there are $m$ candidate sensors in sensor set $\mathcal{S}^{(m)}$, we repeat the above steps $m$ times, and obtain $m$ $\lambda_{i}$ values. Among these $m$ minimum values, we identify the subset that gives the highest $\lambda$ value. The sensor that is included in the subset provides the maximum worst-case observability, that is, $i^*=\max_i\lambda_{i}$. The indicator vector $\bfe$ is updated. If the cost of extra sensors satisfies the sensor budget constraint $G_e$, the sensor $i^*$ is then added to the selected set to form $\mathcal{S}_{i^*}$, which contains $q+1$ sensors. The above procedure can be carried out to add more sensors one by one until the sensor budget constraint is not satisfied. The updated $\mathcal{S}_{res}$ is the solution. Algorithm~\ref{robust set} summarizes the above discussed resilient sensor selection procedures.


\begin{algorithm}[t]
\label{robust set}
\caption{Resilient sensor selection algorithm based on adding sensors one by one}
	\KwIn{The selected set $\mathcal{S}_{opt}$ in Algorithm 1, all sensor set: $\mathcal{S}^{(m)}$, the number of sensors to be added: $w$, the resilient budget constraint $G_{res}:=G+G_e$ }
	\For{k=1:w}{
		\For{$i\in\mathcal{S}^{(m)}$}{
			$\mathcal{S}_{i}=\mathcal{S}_{opt}\cup \{i\}$\\
			\For{$j\in\mathcal{S}_{i}$}{
		        $\mathcal{S}_{ij}$=$\mathcal{S}_{i} \backslash \{j\}$\\
		        Calculate $\lambda_{ij}(\bfz+\bfe-\bfmu)$ based on $\mathcal{S}_{ij}$}
            Choose the worst $\lambda_{i}(\bfz+\bfe-\bfmu)=\min_j\lambda_{ij}(\bfz+\bfe-\bfmu)$
        }
        New added $i^*=\max_i\lambda_{i}(\bfz+\bfe-\bfmu)$\\
        Update the indicator vector $\bfz$\\
        \If{$\bfg^{\rm T}\bfe<G_e$}{$\mathcal{S}_{opt} = \mathcal{S}_{i^*}$ }
        \Else{Terminate the procedure}
        }		
	\KwOut{The resilient sensor set $\mathcal{S}_{res}=\mathcal{S}_{opt}$}
\end{algorithm} 
  
\begin{remark}
	We define $\mathcal{S}_{del}:=\mathcal{S}^{(m)} \backslash \mathcal{S}_{opt}$ as the set of removed sensors obtained by Algorithm 1. However, in Algorithm 2, the newly added sensor that enhances system robustness is selected from $\mathcal{S}^{(m)}$ instead of $\mathcal{S}_{del}$. In the case of considering sensor malfunction and the sensor set being more robust, it makes practical sense to duplicate sensor already in $\mathcal{S}_{opt}$, since some sensors may be crucial for system observability. The effectiveness of the above analysis will also be illustrated through simulations in Section 6.
\end{remark}

\begin{remark}
	The computational complexity of one-by-one removal part for resilient sensor selection is $\mathcal{O}(mr)$. The total computational complexity for resilient sensor selection remains $\mathcal{O}(m^2)$.
\end{remark}

Once the selected set of sensors has been determined, state estimation can be performed using existing estimation methods, such as those presented in \cite{Yin_TSMCS2022,DF2022_Auto_LiuSY,DF2022_SPL_ZhangX}. For the PCCP, preliminary results for state estimation have been reported in \cite{Yin2020_ACSP} for a single absorber of the PCCP. However, in this work, we consider state estimation for the entire PCCP, which involves a larger number of states and algebraic states. To handle state estimation in a computationally efficient manner, we choose to use extended Kalman filtering. The corresponding results will be presented in the next section.

\section{Simulation results}
\label{Examples}

In this section, we apply the proposed algorithms to determine the (near-)optimal sensor set for the PCCP. As shown in Figure \ref{lsy11_Sensorplacement}, there are 23 candidate sensors for potential selection for the PCCP. Each candidate sensor can provide real-time measurements for one state. The 23 candidate sensors shown in Table~\ref{lsy11_taba_sensors} form the initial sensor set $\mathcal{S}^{(23)}=\{C_1, \dots, C_5, T_1, \ldots, T_5, C_6, \ldots, C_{10}, T_6, \ldots, T_{13}\}$. The cost of each candidate sensor is described by the following vector: $\bfg=[20, 20, 20, 20, 20, 1, 1, 1, 1,\notag\\ 1, 20, 20, 20, 20, 20, 1, 1, 1, 1, 1, 1, 1,1]^{\rm T}$, which refers the cost of concentration and temperature sensors in Section \ref{Sensor_cost}. The budget constraint $G$ is set to 72. The dynamic model of the PCCP is discretized at a sample interval $\vartriangle=2$ minutes, and the sensitivity matrix in (\ref{lsy11_4.1c}) is calculated at each sampling point. We assume that the PCCP is affected by process noise that follows a Gaussian distribution of zero mean and standard deviation of $0.4\%$ of the steady-state point corresponding to constant inputs $\bfu=[F_L, Q_{reb}, F_G]=[0.5812{\rm L/s}, 0.1357{\rm KJ/s}, 1{\rm m^3/s}]$, and measurement noise that follows a Gaussian distribution of zero mean and standard deviation of $0.02$ times output vector.

\begin{figure}[!hbt]
	\centering
	\includegraphics[width=\hsize]{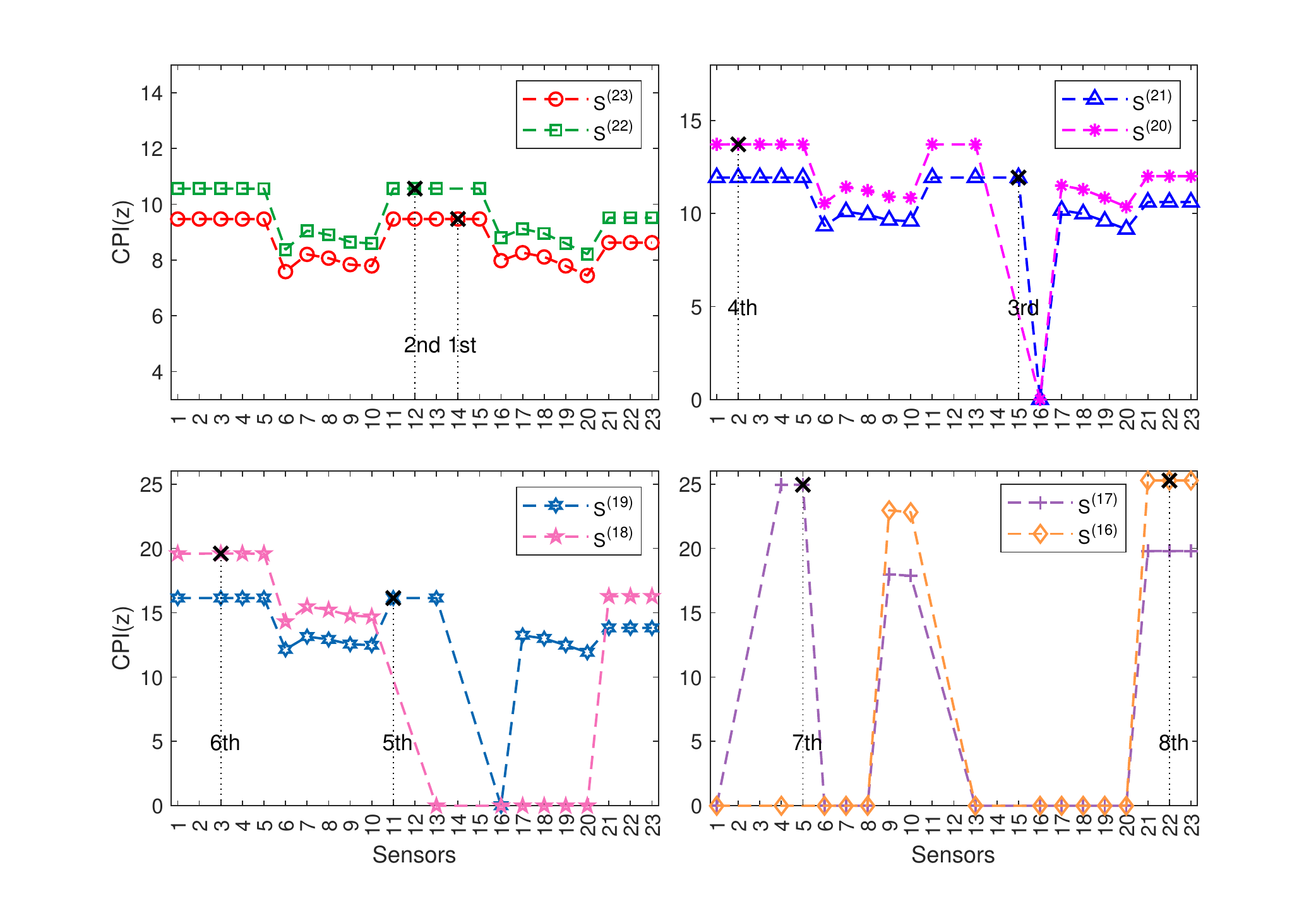}
	\caption{CPI$(\bfz)$ based on removal of different sensors from remaining sensor set.}
	\label{lsy11_fig_CPI}
\end{figure}

Algorithm 1 selects sensors one by one from the initial set $\mathcal{S}^{(23)}$ (expressed as $\mathcal{S}^{(23)}=\{1,2,\dots,23\}$) to find the optimal sensor set and ensure the observability of the system. The largest $\lambda(\bfz)$ value is observed with $\mathcal{S}^{(23)}$, which is $\lambda_{\max}(\bfz)=1828.94$. The initial CPI$(\bfz)=8.59$ is obtained at the total cost $\sum_{i=1}^{23}g_i=213$. The corresponding CPI$(\bfz)$ of the remaining sensor combinations are shown in Figure~\ref{lsy11_fig_CPI} and the steps are recorded in Table~\ref{lsy11_tab} ($m$ is the number of sensors in sensor set, $\mathcal{U}$ presents the set of removed sensors, $c_{actual}$ denotes the current cost of the selected sensors). 

Figure~\ref{lsy11_fig_CPI} shows the corresponding CPI$(\bfz)$ for each sensor combination obtained by removing one sensor from the remaining sensor set. For $\mathcal{S}^{(23)}$, there are 23 red circles in the figure, each of which means the CPI$(\bfz)$ value corresponding to the sensor set containing 22 sensors after removing the sensor with the corresponding number. We select the sensor with the highest CPI$(\bfz)$ value and remove it from $\mathcal{S}^{(23)}$. It is observed that there are some similar values in the figure, such as the values corresponding to the 11th to 15th red circles. Although the size of some values looks similar in the figure, they are not equal (these 5 values are 9.47057, 9.47152, 9.47148, 9.47221, 9.47204, respectively).  The removal of the 14th sensor, which measures $x_{84}$ ($C_{4}^G$(CO$_2$)) of the desorber, leads to the largest increase in CPI$(\bfz)$, raising it to 9.47221. In the next step, we focus on the reduced sensor set $\mathcal{S}^{(22)}$, there are 22 green squares in Figure~\ref{lsy11_fig_CPI}, and the position corresponding to 14th sensor has no green square because the 14th sensor has been removed. According to the values corresponding to these 22 green squares, it is found that if the 12th one is removed from $\mathcal{S}^{(22)}$, the CPI$(\bfz)$ is the largest (10.56) among all possible combinations of 21 sensors. Therefore, the 12th sensor is removed at this step. From identifying the third sensor that should be removed, there will be some CPI$(\bfz)=0$ in the figure, which means that the removal of a certain sensor makes the system unobservable (Equation~(\ref{lsy11_4.2g1}) does not hold). The remaining steps for removing sensors are similar until eight sensors are removed, the cost of the remaining sensors is 72 that meets the budget and system observability. Specifically, sensors $\{14,12,15,2,11,3,5,22\}$ are removed sequentially. Finally, fifteen sensors are selected with CPI$(\bfz)=25.28$. If the budget $G$ for example, is set to 80, then the selection process will terminate when the first seven sensors are removed, corresponding to a cost of 73 to satisfy the budget constraint. If we observe the $\lambda(\bfz)$ value in the Table~\ref{lsy11_tab}, we can find it is gradually reduced as the number of sensors decreases.

\begin{table}[!hbt] 
	\centering
	\caption{The process of sensor selection based on CPI$(\bfz)$ for PCCP.}
	\label{lsy11_tab}
	\renewcommand{\arraystretch}{1.1}
	\tabcolsep 14pt
	\begin{tabular}{cccccc}\hline
 $m$ & $\mathcal{U}$ & $\lambda(\bfz)$ & $c_{actual}$ & CPI$(\bfz)$ & Removed sensor \\\hline
 23  & $\{\}$ & 1828.94 & 213 & 8.59  & 14    \\
 22  & $\{14\}$  & 1828.14 & 193 & 9.47  &  12   \\
 21  & $\{14,12\}$  & 1827.19 & 173 & 10.56 &  15 \\
 20  & $\{14,12,15\}$  & 1826.21 & 153 & 11.94 & 2 \\
 19  & $\{14,12,15,2\}$   & 1825.14 & 133 & 13.72 & 11 \\
 18  & $\{14,12,15,2,11\}$  & 1823.83 & 113 & 16.14 & 3 \\
 17  & $\{14,12,15,2,11,3\}$   & 1822.47 & 93  & 19.60 & 5 \\
 16  & $\{14,12,15,2,11,3,5\}$   & 1820.88 & 73  & 24.94 & 22\\
 15  & $\{14,12,15,2,11,3,5,22\}$  & 1819.88 & 72  & 25.28 & \\\hline
\end{tabular}\end{table}

Table~\ref{lsy11_taba1} shows the selected sensors and cost savings when the budget constraint is set to 72 in Algorithm 1. From the table, 12 temperature sensors and 3 concentration sensors are selected from a pool of 23 sensors. As a result of this selection, the sensor cost is reduced from 213,000 USD to 72,000 USD, representing a 66.2$\%$ decrease in cost. Algorithm 1 can provide the optimal sensor selection based on the degree of observability. 

\begin{table}[!hbt] 
	\centering
	\caption{The selected subset obtained from the proposed algorithm and cost saving. }
	\label{lsy11_taba1}
	\renewcommand{\arraystretch}{1.1}
	\tabcolsep 1pt
	\begin{tabular}{c|c|c|c|c}\hline
		Sensors & Unit price & Total number & Selected sensor set& Cost saving $(\%)$\\\hline
		Temperature          & 1000   & 13 &  $\{6,7,8,9,10,16,17,18,19,20,21,23\}$ &  1000\\\hline
		CO$_2$ concentration & 20,000 & 10 & $\{1,4,13\}$   &  140,000\\\hline
		Total cost & \multicolumn{2}{c|}{213,000} & 72,000  &  141,000 (66.2$\%$)\\\hline
	\end{tabular}
\end{table}

Table~\ref{lsy11_taba3} presents a detailed comparison of the sensor selection results for three different budgets based on $\lambda(\bfz)$. With the increase in the budget, $\lambda(\bfz)$ corresponding to the recommended set of sensors increases.

\begin{table}[!hbt] 
	\centering
	\caption{Sensor selection based on degree of observability $\lambda(\bfz)$ for different budgets. }
	\label{lsy11_taba3}
	\renewcommand{\arraystretch}{1.1}
	\tabcolsep 19pt
	\begin{tabular}{cccc}\hline
		$G$ & Temperature sensors & Concentration sensors & $\lambda(\bfz)$  \\\hline
		91  & $\{6,7,8,9,10,16,17,18,19,20,23\}$ & $\{1,4,11,13\}$ &  1820.18 \\
		111 & $\{6,7,8,9,10,16,17,18,19,20,23\}$ & $\{1,4,5,11,13\}$ & 1821.78 \\
		131 & $\{6,7,8,9,10,16,17,18,19,20,23\}$ & $\{1,3,4,5,11,13\}$ & 1823.14 \\\hline
	\end{tabular}
\end{table}

\begin{figure}[!hbt]
	\centering
	\includegraphics[width=\hsize]{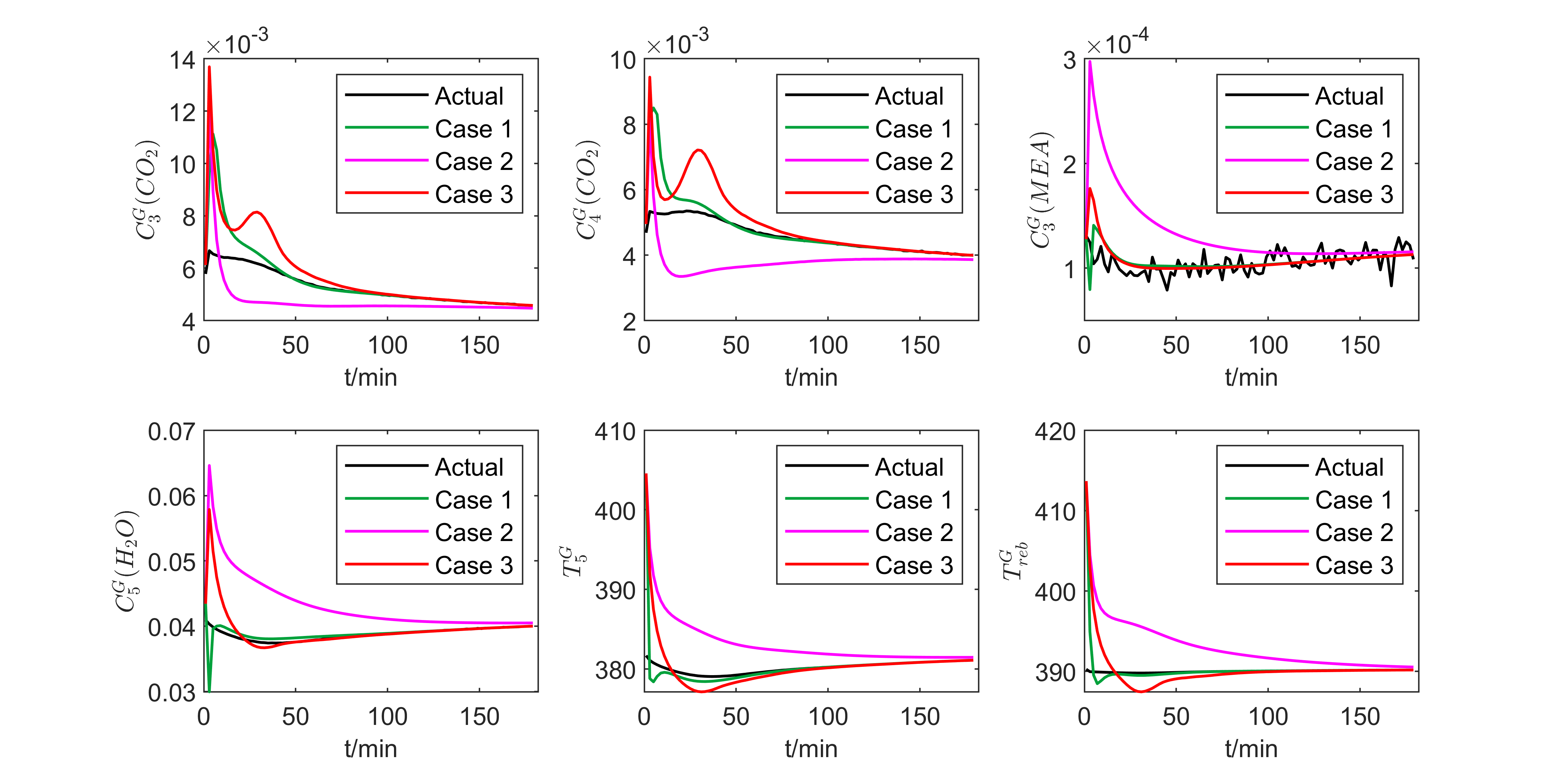}
	\caption{The actual states and their estimates based on different sensor sets (Case 1: Selected by proposed method; Cases 2 and 3: Randomly selected).}
	\label{lsy11_fig_stateestimation}
\end{figure}

After the selected subset is determined, we develop state estimation using the extended Kalman filtering (EKF) to verify the effectiveness of the selected sensors. The initial guess is set to be $1.1\bfx_s$ ($\bfx_s$ is the steady-state value), and the tuning parameters used in the EKF are $\bfQ_w={\rm diag}((0.004\bfx_s)^2)$, $\bfR_v={\rm diag}((0.02\bfy_s)^2)$ and $\bfP(0)={\rm diag}((0.01\bfx_s)^2)$. Some of the actual states and their state estimates are shown in Figure~\ref{lsy11_fig_stateestimation} (Case 1). The state estimates can effectively track the actual state trajectory. To illustrate the superiority of the proposed sensor selection method, we randomly generate two sensor combinations, each of which contains 15 sensors (i.e., Case 2 and Case 3), and the corresponding state estimation results are also presented in Figure~\ref{lsy11_fig_stateestimation}. The state estimation performance of Case 1 is better than that of Cases 2 and 3 which use randomly selected sensors 

\begin{figure}[!hbt]
	\centering
	\includegraphics[width=\hsize]{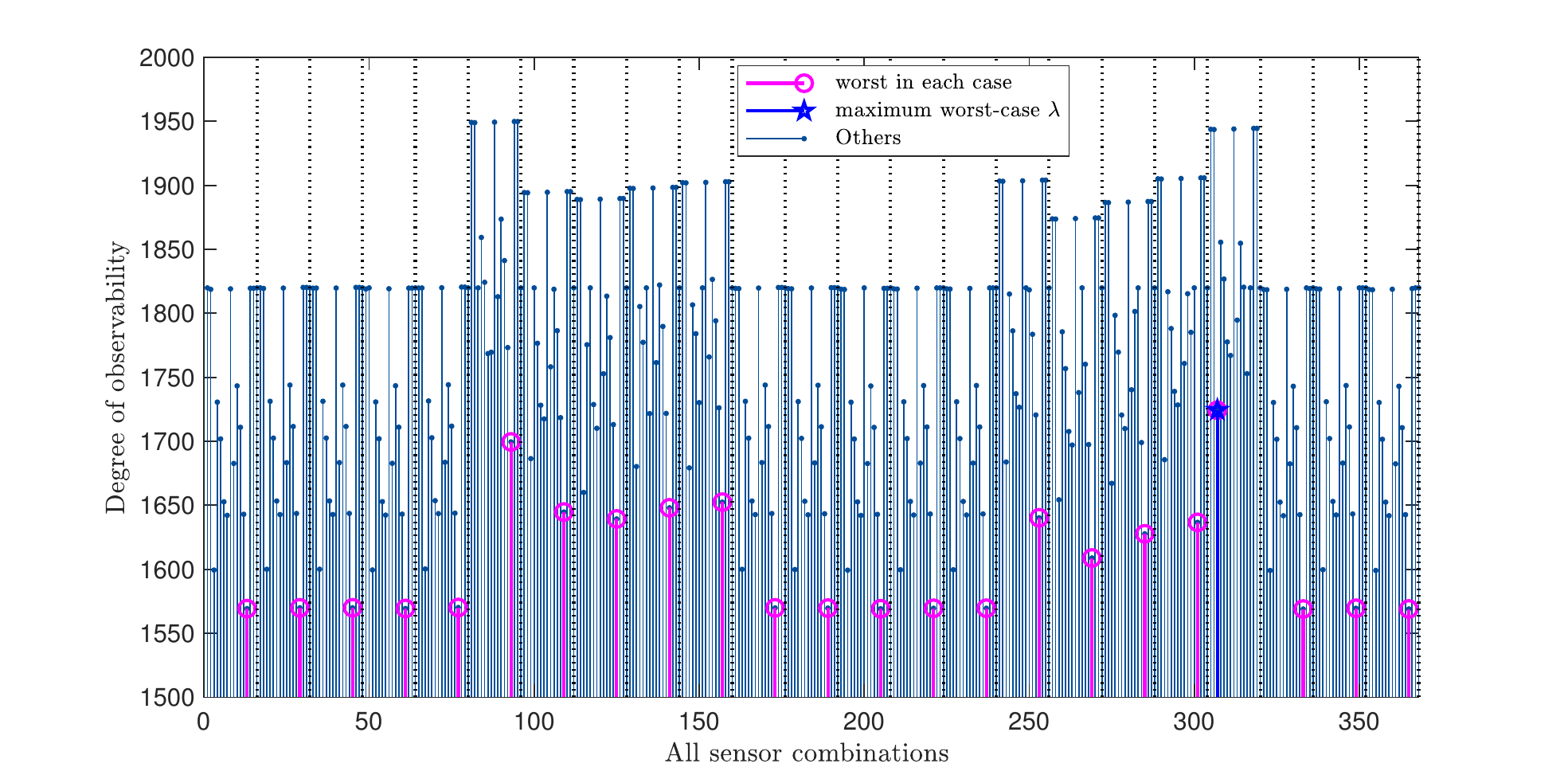}
	\caption{Degrees of observability for all 368 sensor combinations by using the resilient sensor selection algorithm (One sensor fails randomly. Select the first sensor to add to set).}
	\label{lsy11_fig_robust1}
\end{figure}

Next, we conduct additional simulations to demonstrate how the resilient sensor selection algorithm can be used to enhance the robustness of the selected sensors against random sensor malfunction. Starting from the solution obtained by Algorithm 1, we employ Algorithm 2 to identify the most suitable sensor to add and make the newly formed set more robust. Here, we consider an extra budget constraint $G_e$, which allows adding a maximum of $w=1$ sensor. 

First, one sensor is arbitrarily selected from $\mathcal{S}^{(m)}$, which gives a total of $C_{23}^1=23$ cases for the first step. The chosen sensor is then added to $\mathcal{S}{opt}$, resulting in an updated set $\mathcal{S}i=\mathcal{S}{opt}\cup {i}$ containing 16 sensors. Assuming that one of sensors in $\mathcal{S}_i$ malfunctions, there are $C_{16}^1=16$ cases in total. Therefore, we need to detect the $\lambda$ values of $C_{23}^1C_{16}^1=368$ sensor combinations to determine the most suitable sensor to be added. Figure~\ref{lsy11_fig_robust1} shows the process of determining the sensor by testing 368 sensor combinations. These 368 $\lambda$ values are divided into 23 groups, with each group containing 16 $\lambda$ values. These 16 values in the first group, for example, indicates the effect of adding sensor $i=1\in\mathcal{S}^{(23)}$ to $\mathcal{S}{opt}$. Assuming that the sensors in the new set malfunction in turn, the corresponding $\lambda$ values are calculated. We then identify the worst value among the 16 and mark it with a circle in Figure~\ref{lsy11_fig_robust1}. Similarly, the second group represents the effect of adding sensor $i=2\in\mathcal{S}^{(23)}$ to $\mathcal{S}_{opt}$, computing $\lambda$ values, and selecting the worst one to mark. In this way, we identify the 23 worst values, and compare them to choose the largest one, which corresponds to the 20th $\lambda$ marked in Figure~\ref{lsy11_fig_robust1}. The sensor label associated with the 20th $\lambda$ value is the optimal result obtained when random malfunction is considered, thus, sensor $i=20$ should be added to $\mathcal{S}_{opt}$.

\begin{figure}[!hbt]
	\centering
	\includegraphics[width=\hsize]{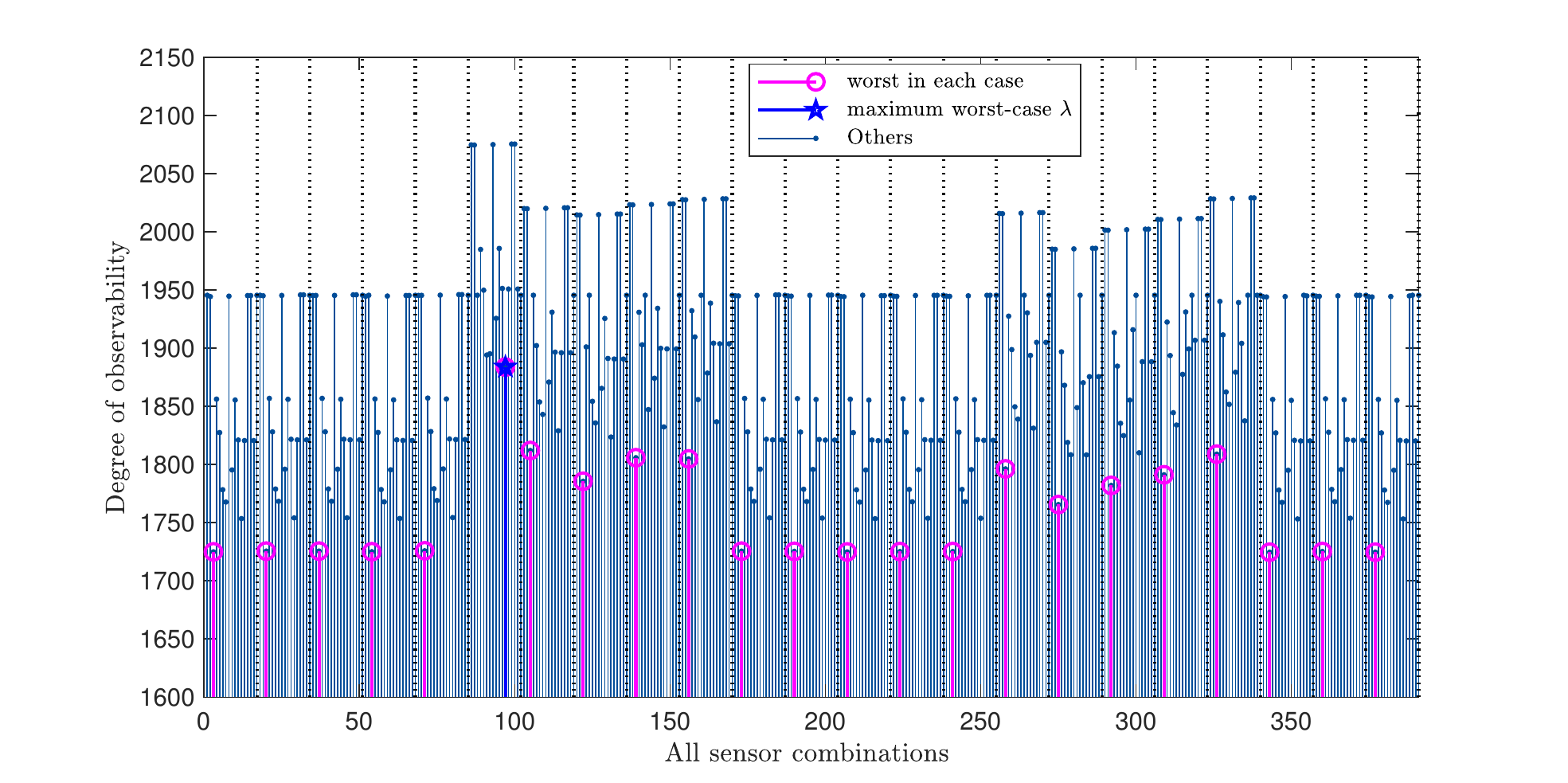}
	\caption{Degrees of observability for all 391 sensor combinations by using the resilient sensor selection algorithm (One sensor fails randomly. Select the second sensor to add to set).}
	\label{lsy11_fig_robust2}
\end{figure}

The updated resilient sensor set is $\mathcal{S}_{res}=\mathcal{S}_{opt} \cup \{20\}$, which now contains 16 sensors. If the extra budget constraint $G_e$ allowed for selecting a second sensor ($w=2$) to further increase the robustness of the sensor set, we needed to consider random malfunction of 17 sensors ($16+1$). Therefore, for each sensor $i\in\mathcal{S}^{(23)}$ that we considered adding to the resilient sensor set, 17 values of $\lambda$ need to be calculated, resulting in $C_{23}^1C_{17}^1=391$ sensor combinations to test. Figure~\ref{lsy11_fig_robust2} illustrates the process of determining the second suitable sensor added to the set $\mathcal{S}_{res}$ by testing 391 sensor combinations. Similarly, we follow the same procedure as before, selecting the worst value among the 17 in each of the 23 groups and then identifying the largest of the 23 marked $\lambda$. Then, the 6th sensor is selected to add to $\mathcal{S}_{res}$. Figures~\ref{lsy11_fig_robust1} and \ref{lsy11_fig_robust2} together demonstrate the selection of the most suitable two sensors to add to the set $\mathcal{S}_{res}$, making the updated set more robust when considering random malfunction of one sensor.

\begin{figure}[!hbt]
	\centering
	\includegraphics[width=\hsize]{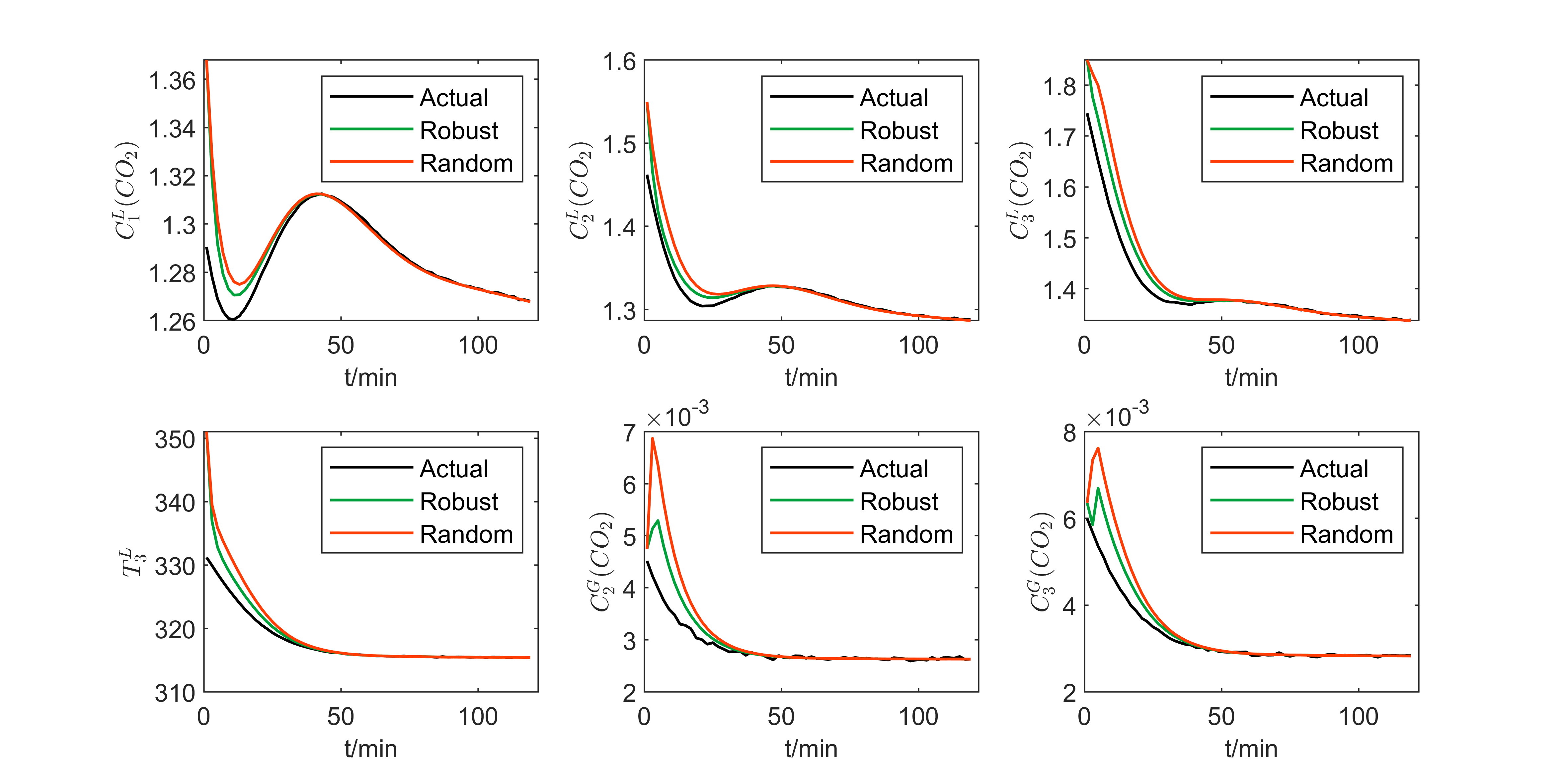}
	\caption{The actual states and their estimates based on different sensor sets.}
	\label{lsy11_fig_state_robust}
\end{figure}

According to the results of resilient sensor selection algorithm, we identify the two sensors to be added: sensors 20 and 6. In the following simulation, we randomly select two sensors as the new ones. For two different sets, the same two sensors malfunction during state estimation using the EKF to verify the effectiveness of the resilient sensor set. The initial guess is set to be the same as the former. Figure~\ref{lsy11_fig_state_robust} shows some of the actual states and their state estimates obtained using the two different sets. It is obvious that the state estimates obtained by resilient sensor set can track the actual state trajectory better than that of the randomly selected set.

\section{Conclusions}
\label{Conclusions}

In this work, In this study, we have developed a sensitivity-based method to quantify the degree of observability of post-combustion CO$_2$ capture plants. We also propose a computationally efficient optimal sensor selection algorithm that can determine (near)-optimal sensor locations by maximizing the cost performance index under different budget constraints. Furthermore, we introduced a resilient sensor selection algorithm that enhances the robustness of the sensor set in the presence of random sensor malfunction. Through the proposed algorithms, we have improved the economy, reduced the complexity, and increased the robustness of the sensor selection problem for the post-combustion CO$_2$ capture plants. 

\section{Acknowledgement}

The first author, S.Y. Liu, is a visiting Ph.D. student in the Department of Chemical and Materials Engineering at the University of Alberta from March 2021 to February 2023. She acknowledges the financial support from the China Scholarship Council (CSC) during this period.



\begin{thebibliography}{10}

\bibitem{MacDowell2010_RSC}
N.~MacDowell, N.~Florin, A.~Buchard, J.~Hallett, A.~Galindo, G.~Jackson, C.~S.
  Adjiman, C.~K. Williams, N.~Shah, and P.~Fennell, ``An overview of {CO}$_2$
  capture technologies,'' {\em Energy \& Environmental Science}, vol.~3,
  no.~11, pp.~1645--1669, 2010.

\bibitem{Manaf2019_JPC}
N.~A. Manaf, A.~Qadir, and A.~Abbas, ``Efficient energy management of {CO}$_2$
  capture plant using control-based optimization approach under plant and
  market uncertainties,'' {\em Journal of Process Control}, vol.~74, pp.~2--12,
  2019.

\bibitem{Zhang2015_TAC}
H.~Zhang, P.~Cheng, L.~Shi, and J.~Chen, ``Optimal denial-of-service attack
  scheduling with energy constraint,'' {\em IEEE Transactions on Automatic
  Control}, vol.~60, no.~11, pp.~3023--3028, 2015.

\bibitem{Mo2010_IEEECDC}
Y.~Mo, E.~Garone, A.~Casavola, and B.~Sinopoli, ``False data injection attacks
  against state estimation in wireless sensor networks,'' in {\em 49th IEEE
  Conference on Decision and Control (CDC)}, pp.~5967--5972, IEEE, 2010.

\bibitem{Ye2020_TAC}
L.~Ye, N.~Woodford, S.~Roy, and S.~Sundaram, ``On the complexity and
  approximability of optimal sensor selection and attack for {K}alman
  filtering,'' {\em IEEE Transactions on Automatic Control}, vol.~66, no.~5,
  pp.~2146--2161, 2020.

\bibitem{Ye2020_IEEETCNS}
L.~Ye, S.~Roy, and S.~Sundaram, ``Resilient sensor placement for {K}alman
  filtering in networked systems: Complexity and algorithms,'' {\em IEEE
  Transactions on Control of Network Systems}, vol.~7, no.~4, pp.~1870--1881,
  2020.

\bibitem{LiuSY2022_ADCONIP}
S.~Y. Liu, X.~Yin, and J.~Liu, ``Sensor placement for wastewater treatment
  plants: a computationally efficient algorithm,'' {\em 7th International
  Symposium on Advanced Control of Industrial Processes}, pp.~1--6, 2022.

\bibitem{Shastri2006_WRPM}
Y.~Shastri and U.~Diwekar, ``Sensor placement in water networks: A stochastic
  programming approach,'' {\em Journal of water resources planning and
  management}, vol.~132, no.~3, pp.~192--203, 2006.

\bibitem{Rrico2007_CCE}
V.~Rico-Ramirez, S.~Frausto-Hernandez, U.~M. Diwekar, and S.~Hernandez-Castro,
  ``Water networks security: A two-stage mixed-integer stochastic program for
  sensor placement under uncertainty,'' {\em Computers \& chemical
  engineering}, vol.~31, no.~5-6, pp.~565--573, 2007.

\bibitem{Mkwananzi2022_JPC}
T.~Mkwananzi, T.~M. Louw, L.~Auret, M.~Mandegari, and J.~F. G{\"o}rgens,
  ``Combined optimal sensor network design and self-optimizing control with
  application in a typical sugarcane mill,'' {\em Journal of Process Control},
  vol.~114, pp.~82--91, 2022.

\bibitem{Zhang2017_Auto}
H.~Zhang, R.~Ayoub, and S.~Sundaram, ``Sensor selection for {K}alman filtering
  of linear dynamical systems: Complexity, limitations and greedy algorithms,''
  {\em Automatica}, vol.~78, pp.~202--210, 2017.

\bibitem{Alonso2004_CCE}
A.~A. Alonso, I.~G. Kevrekidis, J.~R. Banga, and C.~E. Frouzakis, ``Optimal
  sensor location and reduced order observer design for distributed process
  systems,'' {\em Computers \& chemical engineering}, vol.~28, no.~1-2,
  pp.~27--35, 2004.

\bibitem{Clark2021_IEEESJ}
E.~Clark, S.~L. Brunton, and J.~N. Kutz, ``Multi-fidelity sensor selection:
  Greedy algorithms to place cheap and expensive sensors with cost
  constraints,'' {\em IEEE Sensors Journal}, vol.~21, no.~1, pp.~600--611,
  2021.

\bibitem{Yamada2021_MSSP}
K.~Yamada, Y.~Saito, K.~Nankai, T.~Nonomura, K.~Asai, and D.~Tsubakino, ``Fast
  greedy optimization of sensor selection in measurement with correlated
  noise,'' {\em Mechanical Systems and Signal Processing}, vol.~158, p.~107619,
  2021.

\bibitem{Jawaid2015_Auto}
S.~T. Jawaid and S.~L. Smith, ``Submodularity and greedy algorithms in sensor
  scheduling for linear dynamical systems,'' {\em Automatica}, vol.~61,
  pp.~282--288, 2015.

\bibitem{Yin2018_CCE}
X.~Yin and J.~Liu, ``State estimation of wastewater treatment plants based on
  model approximation,'' {\em Computers \& Chemical Engineering}, vol.~111,
  pp.~79--91, 2018.

\bibitem{Sahoo2019_AIChE}
S.~R. Sahoo, X.~Yin, and J.~Liu, ``Optimal sensor placement for
  agro‐hydrological systems,'' {\em AIChE Journal}, vol.~65, no.~12, 2019.

\bibitem{Singh2006_IECR}
A.~K. Singh and J.~Hahn, ``Sensor location for stable nonlinear dynamic
  systems: Multiple sensor case,'' {\em Industrial \& Engineering Chemistry
  Research}, vol.~45, no.~10, pp.~3615--3623, 2006.

\bibitem{Qi2015_IEEETPS}
J.~Qi, K.~Sun, and W.~Kang, ``Optimal {PMU} placement for power system dynamic
  state estimation by using empirical observability {G}ramian,'' {\em IEEE
  Transactions on Power Systems}, vol.~30, no.~4, pp.~2041--2054, 2015.

\bibitem{Sumana2009_JPC}
C.~Sumana and C.~Venkateswarlu, ``Optimal selection of sensors for state
  estimation in a reactive distillation process,'' {\em Journal of Process
  Control}, vol.~19, no.~6, pp.~1024--1035, 2009.

\bibitem{Awasthi2020_JPC}
U.~Awasthi, K.~A. Palmer, and G.~M. Bollas, ``Optimal test and sensor selection
  for active fault diagnosis using integer programming,'' {\em Journal of
  Process Control}, vol.~92, pp.~202--211, 2020.

\bibitem{Guo2021_JPC}
X.-G. Guo, P.-Y. Hong, and T.-M. Laleg-Kirati, ``Calibration and validation for
  a real-time membrane bioreactor: A sliding window approach,'' {\em Journal of
  Process Control}, vol.~98, pp.~92--105, 2021.

\bibitem{LiuSY2022_DYCOPS}
S.~Liu, X.~Yin, and J.~Liu, ``Simultaneous state and parameter estimation of
  not fully measured systems: a distributed approach,'' {\em
  IFAC-PapersOnLine}, vol.~55, no.~7, pp.~1--6, 2022.

\bibitem{Decardi-Nelson2018_Process}
B.~Decardi-Nelson, S.~Liu, and J.~Liu, ``Improving flexibility and energy
  efficiency of post-combustion {CO}$_2$ capture plants using economic model
  predictive control,'' {\em Processes}, vol.~6, no.~9, p.~135, 2018.

\bibitem{Harun2012_IJGGC}
N.~Harun, T.~Nittaya, P.~L. Douglas, E.~Croiset, and L.~A. Ricardez-Sandoval,
  ``Dynamic simulation of {MEA} absorption process for {CO}$_2$ capture from
  power plants,'' {\em International Journal of Greenhouse Gas Control},
  vol.~10, pp.~295--309, 2012.

\bibitem{Liu2021_IECR}
J.~Liu, A.~Gnanasekar, Y.~Zhang, S.~Bo, J.~Liu, J.~Hu, and T.~Zou,
  ``Simultaneous state and parameter estimation: the role of sensitivity
  analysis,'' {\em Industrial \& Engineering Chemistry Research}, vol.~60,
  no.~7, pp.~2971--2982, 2021.

\bibitem{LiuSY2022_ChERD}
S.~Y. Liu, X.~Yin, J.~Liu, J.~Liu, and F.~Ding, ``Distributed simultaneous
  state and parameter estimation of nonlinear systems,'' {\em Chemical
  Engineering Research and Design}, vol.~181, pp.~74--86, 2022.

\bibitem{Aissi2009_EJOR}
H.~Aissi, C.~Bazgan, and D.~Vanderpooten, ``Min--max and min--max regret
  versions of combinatorial optimization problems: A survey,'' {\em European
  journal of operational research}, vol.~197, no.~2, pp.~427--438, 2009.

\bibitem{Yin_TSMCS2022}
X.~Yin and B.~Huang, ``Event-triggered distributed moving horizon state
  estimation of linear systems,'' {\em IEEE Transactions on Systems, Man, and
  Cybernetics: Systems}, vol.~52, no.~10, pp.~6439--6451, 2022.

\bibitem{DF2022_Auto_LiuSY}
S.~Liu, X.~Zhang, L.~Xu, and F.~Ding, ``Expectation--maximization algorithm for
  bilinear systems by using the {R}auch--{T}ung--{S}triebel smoother,'' {\em
  Automatica}, vol.~142, p.~110365, 2022.

\bibitem{DF2022_SPL_ZhangX}
X.~Zhang and F.~Ding, ``Optimal adaptive filtering algorithm by using the
  fractional-order derivative,'' {\em IEEE Signal Processing Letters}, vol.~29,
  pp.~399--403, 2022.

\bibitem{Yin2020_ACSP}
X.~Yin, B.~Decardi-Nelson, and J.~Liu, ``Distributed monitoring of the
  absorption column of a post-combustion {CO}$_2$ capture plant,'' {\em
  International Journal of Adaptive Control and Signal Processing}, vol.~34,
  no.~6, pp.~757--776, 2020.

\end{thebibliography}

\end{document}